\documentclass{aa}

\usepackage{txfonts}
\usepackage{natbib}
\usepackage{graphicx}
\usepackage{supertabular}
\usepackage{longtable}

\newcommand{\ms}{\mbox{m s$^{-1}~$}}

\begin{document}
\bibliographystyle{aa}

\title{A planetary system with gas giants and super-Earths around the nearby M dwarf GJ 676A}

\subtitle{Optimizing data analysis techniques for the
detection of multi-planetary systems}

\author{
Guillem Anglada-Escud\'e$^{\star}$\inst{1}, 
Mikko Tuomi\thanks{Both authors contributed equally to this work}\inst{2,3}
}

\institute{
       Universit\"{a}t G\"{o}ttingen,
       Institut f\"ur Astrophysik,
       Friedrich-Hund-Platz 1,
       37077 G\"{o}ttingen, Germany.
       \email{guillem.anglada@gmail.com}
       \and
       University of Hertfordshire, Centre for Astrophysics Research, Science
       and Technology Research Institute, College Lane, AL10 9AB, Hatfield, UK.
       \email{m.tuomi@herts.ac.uk}
       \and
       University of Turku, Tuorla Observatory, 
       Department of Physics and Astronomy, 
       V\"ais\"al\"antie 20, FI-21500, Piikki\"o, Finland. 
       \email{miptuom@utu.fi}
}

\date{submitted March 2012}

\begin{abstract}
{
Several M dwarfs are targets of systematical monitoring in searches for Doppler
signals caused by low-mass exoplanet companions. As a result, an emerging
population of high-multiplicity planetary systems around low-mass stars are
being detected as well.
} {
We optimize classic data analysis methods and develop new ones
to enhance the sensitivity towards lower amplitude planets in
high-multiplicity systems. We apply these methods to the public
HARPS observations of GJ 676A, a nearby and relatively quiet M
dwarf with one reported gas giant companion. 
} {
We rederived Doppler measurements from public HARPS spectra 
using the recently developed template matching method 
(HARPS-TERRA software). We used refined versions of periodograms
to assess the presence of additional low-mass companions. We
also analysed the same dataset with Bayesian statistics tools
and compared the performance of both approaches.
} {
We confirm the already reported massive gas giant candidate and
a long period trend in the Doppler
measurements. In addition to that, we find very secure evidence in favour
of two new candidates in close-in orbits and masses in the
super-Earth mass regime. Also, the increased time-span of the
observations allows the detection of curvature in the long-period 
trend. suggesting the presence of a massive outer
companion whose nature is still unclear.
} {
Despite the increased sensitivity of our new periodogram
tools, we find that Bayesian methods are significantly more
sensitive and reliable in the early detection of candidate
signals, but more works is needed to quantify their
robustness against false positives. While hardware development
is important in increasing the Doppler precision, development
of data analysis techniques can help to reveal new results
from existing data sets with significantly fewer resources. This
new system holds the record of minimum-mass range
(from M$\sin i \sim$4.5 M$_\oplus$ to 5 M$_{jup}$) and period range (from
P$\sim$3.6 days to more than 10 years). Although all the planet
candidates are substantially more massive, it is the 
first exoplanetary system with a general architecture similar 
to our solar system. GJ 676A can be happily added to the family of 
high-multiplicity planetary systems around M dwarfs.
}

\end{abstract}


\titlerunning{A planetary system around GJ 676A}
\authorrunning{Guillem Anglada-Escud\'e \& Mikko Tuomi}

\maketitle


\section{Introduction}

Doppler spectroscopy is currently the most effective method for
detecting planet candidates orbiting nearby stars. The current
precision enables the detection of planets of a few Earth
masses in close-in orbits, especially around low-mass
stars \citep[e.g.,][]{mayor:2009}. Two methods are currently used to obtain
precision Doppler measurements in the visible part of the
spectrum: the gas cell and the stabilized spectrograph
approach. The gas cell method consists on inserting a cell
containing iodine gas in the beam of the telescope which
provides a very accurate method to solve for the wavelength
solution, instrumental profile variability, and the Doppler
changes in the stellar spectrum \citep{butler:1996}. The
second approach is based on building a  mechanically stable
fiber-fed spectrograph calibrated with an emission
lamp \citep{baranne:1996}. HARPS is the best example of a
stabilized spectrograph in operation. It is installed at
the 3.6m Telescope in La Silla-ESO observatory/Chile
\citep{harps:construction}. The list of planets detected by
HARPS is long and varied, as can be seen in the 35 papers of the
series \textit{The HARPS search for southern extra-solar
planets}. Instead of citing all of them, we refer the
interested reader to the latest HARPS results presented in
\citet{pepe:2011}, \citet{mayor:2011}, and \citet{bonfils:2011}. 
HARPS has demonstrated a radial velocity (RV) stability at the level of
1 \ms on time-scales of several years. Since January 2011,
reduced data products derived from the HARPS data reduction
software (DRS) are publicly available through a dedicated
webpage in ESO
\footnote{\texttt{http://archive.eso.org/wdb/wdb/eso/repro/form}}.
All data used in this work have been obtained from there. 

The increasing demand for higher Doppler precision has
motivated a significant investment in hardware development, and
a number of new stabilized spectrographs are currently under
construction \citep[e.g., ][]{freqcomb}. It is known, however,
that the method employed by the HARPS-DRS to extract RV
measurements (cross-correlation function) is suboptimal in
exploiting the Doppler information in the stellar spectrum
\citep{queloz:1995,pepe:2002}, therefore developments in the data
analysis techniques are also required to reach photon noise
limited precision. We have recently developed a least-squares
template matching method \citep[HARPS-TERRA software,
][]{anglada:2012a, anglada:2012b} that is able to derive
precise RV measurements from HARPS spectra obtaining a
substantial increment of precision on K and M dwarfs. In
\citet{anglada:2012a}, 34 observations on the M0 dwarf \object{GJ 676A}
were used to illustrate the improvement in precision of the
HARPS-TERRA measurements compared to those used in the
discovery paper of the massive planet candidate \object{GJ 676Ab}
\citep{discovery}. Additional observations on this star have
recently been released through the ESO archive, and we applied
HARPS-TERRA to extract new RV measurements of the full set. In
a preliminary analysis using classic periodogram methods, we
found tentative evidence for additional planets in the system.
However, these preliminary detections did not provide
convincing results. Recent developments in the Bayesian
analysis methods of Doppler data \citep[e.g., ][]{tuomi:2012}
indicate that correlation between parameters seriously affect
the sensitivity of periodogram-based methods in detecting
additional low-amplitude signals. Moreover, careful
Bayesian analyses provide increased sensitivity to lower
amplitude signals \citep{tuomi:2012} and seem to be less prone
to false positives than methods based on sequential
periodogram analyses of the residuals only \citep{tuomi:2011}.

In this work, we develop and test data analysis methods for
optimal detection of low-mass companions in multi-planetary
systems and apply them to the HARPS-TERRA measurements of GJ
676A. In Section \ref{sec:methods}, we describe a new
periodogram-based approach (recursive periodogram) and review
the Bayesian analysis tools also developed to deal with
multi-Keplerian fits. Section \ref{sec:star} reviews the stellar
properties of GJ 676A, describes the observations, discusses
periodicities detected in activity indices and describes the
previously detected candidates \citep[one massive gas giant and
a long-period trend, ][]{discovery}. Section \ref{sec:analysis}
analyses the RVs of GJ 676A using these tools. Both
methods (recursive periodograms and Bayesian analyses) agree in
the detection of two additional sub-Neptune/super-Earth mass
candidates in close-in orbits. We also use the opportunity to
test the sensitivity of both detection methods by applying them
to a subset of observations (first 50 epochs). We find that,
while the recursive periodogram approach is able to only spot one of
the additional signals at low confidence, a Bayesian analysis can already
recover the same candidates obtained from
the full set of observations. In section \ref{sec:activity} we 
identify and discuss a
few periodicities in some of the activity indices. Finally, in
Section \ref{sec:conclusions} we place the unique features of
the planetary system around GJ 676A in the context of the
currently known population of exoplanets and provide some
concluding remarks.

\section{Data analysis methods}\label{sec:methods}

\subsection{Recursive periodograms } \label{sec:recursive}

Classic least-squares periodograms and derived methods
\citep{scargle:1982,cumming:2004} consist of adjusting a
sine-wave (equivalent to a circular orbit) to a list of test
periods and plot these periods against some measure of merit.
When k-periodic signals are detected in the data, the
corresponding Keplerian model is subtracted from the data and
a least-squares periodogram is typically applied to the
residuals to assess if there is a k+1-th periodicity left. As
noted by several authors
\citep{anglada:2010,hd10180,tuomi:2012}, non-trivial
correlations between parameters are likely to decrease the
significance of (yet undetected) low-amplitude signals. That
is, as the number of the Keplerian signals in a model
increase, the aliases of previously detected signals and
other non-trivial correlations seriously affect the
distribution of the residuals and, unless the new signal is
very obvious, a periodogram of the residuals will not properly
identify (even completely confuse) the next most likely
periodicity left in the data. 

To account for parameter correlation at the period search
level, we developed a generalized version of the classic
least-squares periodogram optimized for multiplanet detection
that we call \textit{recursive periodogram}. Instead of
adjusting sine-waves to the residuals only, a recursive
periodogram consists of adjusting all the parameters of the
already detected signals together with the signal under
investigation. Even in there are correlations,
candidate  periods will show prominently as long as the new
solution provides a net improvement of the previous global
solution. In our approach and by analogy to previous
least-squares periodograms, a circular orbit (sinusoid) is
always assumed for the proposed new periodicity. When no
previous planets are detected, this is equivalent to the
generalized least-squares periodogram discussed by
\citet{zechmeister:2009}, and is a natural generalization of
the methods discussed by \citet{cumming:2004} to
multi-Keplerian solutions. The graphical representation of the
periodogram is then obtained by plotting the obtained period
for the new planet (X-axis) versus the F-ratio statistic
obtained from the fit (Y-axis). The highest peak in this
representation indicates, in a leasts-squares sense, the most
likely periodic signal present in the data.

As with any other classic least-squares periodogram method, one has
to assess if adding a new signal is justified given the
improvement of the fit. As proposed by \citet{cumming:2004},
we use the F-ratio statistic to quantify the improvement of
the fit of the new model (k+1 planets) compared to the null
hypothesis (k planets). The F-ratio as a function of the test
period is defined as

\begin{eqnarray}
F(P) = \frac{
   (\chi^2_0-\chi^2_P)/(N_{k+1}-N_{k})
}{
\chi^2_P/(N_{obs}-N_{k+1})
}\,,
\end{eqnarray}

\noindent where 
$\chi^2_0$ is the chi-squared statistic
for the model with k-planets (null hypothesis), 
$\chi^2_P$ is the chi-squared statistic at the 
test period $P$, $N_k$ is the number of free parameters 
in the model with k-planets, and $N_{k+1}$ is the number of free
parameters in the model including one more candidate in a
circular orbit at period $P$. For a circular orbit, the number
of additional parameters $N_{k+1}-N_k$ is 2 (amplitude and
phase of a sinusoid). Assuming large $N_{obs}$,
statistical independence of the observations, and Gaussian
errors, $F(P)$ would follow a Fisher F-distribution with
$N_{k+1}-N_{k}$ and $N_{obs}-N_{k+1}$ degrees of freedom. The
cumulative distribution (integral from 0 to the obtained
F-ratio) is then used as the confidence level $c$ at
each $P$ (also called single-frequency confidence level).
Because the period is a strongly nonlinear parameter, each
peak in a periodogram must be treated as an independent
experiment (so-called independent frequencies). Given a
dataset, the number of independent frequencies $M$ can be
approximated by $P_{min} \Delta T $, where $\Delta$ T is the
time baseline of the observations and $P_{min}$ is the
shortest period (highest frequency) under consideration. Given
M, the  analytic false alarm probability is finally derived
as FAP$=1-c^M$.

Since the fully Keplerian problem is very nonlinear, several
iterations at each test period are necessary to ensure
convergence of the solution. A typical recursive periodogram
consists of testing several thousands of such solutions and,
therefore, it is a time-consuming task. As a result, special
care has to be taken in using a robust and numerically
efficient model to predict the observables. We found that 
a slight modification of the
parameterization given by \citet{wright:2009} provided the
best match to our needs. The only change we applied was using
the initial mean anomaly $M_0$ instead of the time of
periastron $T_0$ as a free parameter. These two quantities are
related by $2\pi T_0/P = -M_0$. From this expresion one can
see that the replacement of $T_0$ by $M_0$ eliminates the
non-linear coupling between $T_0$ and $P$. \citet{wright:2009}
also provide the partial derivatives of the observables
(radial velocity) in a numerically efficient representation.
The partial derivative of the RV with respect to $M_0$
(instead of T$_0$) is trivially obtained as minus the partial
derivative of the radial velocity with respect to the mean
anomaly $E$ ($\partial v/\partial M_0 = -\partial v/\partial
E)$. Beyond this change, the adopted model is identical to the
one given in \citet{wright:2009} so, for the sake of brevity,
we do not provide the full description here. To accelerate
convergence at each test period, we first solve for linear
parameters only. Next, we use the
Levenberg-Marquardt \citep{levenberg} method to smoothly
approach the $\chi^2$ minimum and, finally, a few interations
of a straight nonlinear least-squares solver
\citep[][]{numerical} are used to quickly converge to the
final solution. Although fitting for k-planets at
each test period would seem a very time-consuming effort, we
are implicitly assuming that the solver is already close to
the $\chi^2$ local minimum. Therefore, relatively few
nonlinear iterations (between 20 and 50) are typically enough
to reach the closest local minimum. Since all orbits are
re-adjusted, and even though the method still suffers from some
of the typical pitfalls of sequential Keplerian fitting (e.g.,
the solver can still become stuck on local minima), the solution at
each test period always has a higher significance than a
periodogram on the residuals, especially when parameters
are correlated.

It is known that the assumptions required by the F-ratio
tests  might not be stricly satisfied by RV observations.
Therefore, an empirical scheme is always desirable to better
asses the FAP of a new detection. Since the recursive
periodogram is a straight generalization of least-squares
periodograms, we adopted the brute-force Monte Carlo method
proposed by \citet{cumming:2004} to obtain empirical estimates
of the FAPs. That is, we computed recursive periodograms on
synthetic datasets and counted how many times spurious peaks
with higher power than the signal under investigation were
obtained by an unfortunate arrangement of the noise. Each
Monte Carlo trial consists of : 1) taking the residuals to the
model with k-planets and randomly permutate them over the same
observing epochs, 2) adding back the signal of the model with
k-planets, 3) re-adjusting the solution with k-planets (new
null hypothesis), 4) computing the recursive periodogram on
this new synthetic dataset and, 5) recording the highest
F-ratio in a file. The FAP will be the number of times we obtain
an F-ratio higher than the original one divided by the number
of trials.

A recursive periodogram can take a few tens of minutes
depending on the number of datapoints and number of planets in
the model. While this is not a serious problem while exploring
one dataset, it becomes a problem when FAPs have to be
empirically computed for many thousands of Monte Carlo trials.
As a general rule, we accept new candidates if they show an
empirical FAP lower than 1\%. While this threshold is
arbitrary, it guarantees that even if some of the proposed
candidates are false positives, we will not seriously
contaminate the current sample of $\sim$700 RV detected
exoplanets with spurious detections. As a first saving measure
and given that analytic FAPs are known to be over-optimistic,
we only compute empirical FAPs if the analytical FAP
prediction is already lower than $1\%$. The chance of
obtaining a false alarm in a trial is a Poisson process and,
therefore, the uncertainty in the empirical FAP is
$\sim\sqrt{N_{\rm FAPs}}/N_{\rm trials}$. Our aim is to
guarantee that the empirical FAP is $<$ 1\% at a 4-$\sigma$
level, so we designed the following strategy to minimize the
the number of Monte Carlo trials. That is, we first run 1000
trials. If no false-alarms are detected, the candidate is
accepted and the analytic FAP is used to provide an estimate
of the real one. If the number of trials generating false
alarms is between 1 and 20 (estimated FAP$\sim$0.1--3\%), we
extend the number of trials to 10$^4$. If the updated FAP is
lower than $0.5\%$, we stop the simulations and accept the
candidate. If the empirical FAP is still between $0.5$ and
$1.5\%$, $5\,10^4$ trials are obtained and the derived FAP is
used to decide if a candidate is accepted. While the
computation time for 1000 trials in a single processor is
prohibitively high, the computation of many recursive
periodograms can be easily parallelized in modern
multi-processor desktop computers. For the GJ 676A dataset and
a (3+1)-planet model, 10$^3$ trials would take 2.3 days on one
2.0 GHz CPU. The same computation on 40 logical CPUs takes 1.4
hours, allowing one to obtain empirical FAP runs with
10$^4$--10$^5$ trials in less than a week.

\subsection{Bayesian analysis methods} \label{sec:bayesian}

As in e.g. \citet{tuomi:2012}, the Bayesian analyses of the
RVs of GJ 676A were conducted using samplings of the posterior
probability densities, estimation of Bayesian evidences,
and the corresponding model probabilities based on these
samples.

We sampled the posterior densities using the adaptive
Metropolis algorithm \citep{haario:2001}, also described
extensively in \citet{tuomi:2011}. Because it converges reliably
and relatively rapidly to the posterior density in most
situations, we performed several samplings of the parameter
space of each model. Different samplings were started with
different initial states to ensure that the global
probability maximum of the parameter space was found for each
model. If they all converged to the same solution, we could
confidently conclude that the corresponding maximum was indeed
the global one. This check was performed because it is
possible that the Markov chains becomes stuck in a local maximum
if it is sufficiently high and the initial state happens to be
close to it. As a result, we could then reliably estimate the
parameters using the maximum \emph{a posteriori} (MAP)
estimates and Bayesian credibility sets (BCSs) as uncertainty
estimates \citep{tuomi:2009}.

The Bayesian evidence of each model was calculated using the
one block Metropolis-Hastings (OBMH) estimate
\citep{chib:2001}. It requires a statistically representative
sample from the posterior, available due to posterior
samplings, and can be used to assess the evidence and the
corresponding model probabilities with relatively little
computational effort when determining the number of Keplerian
signals in an RV data set favoured by the data
\citep[e.g.][]{tuomi:2011,tuomi:2012,tuomi:2011b}.

Using the OBMH estimates, we determined the probabilities of
the models with differing numbers of Keplerian signals.
However, we did not blindly choose the model with the greatest
posterior probability and added it to the solution unless three
detection criteria were also satisfied. We required that (1)
the posterior probability of a model with $k+1$ Keplerian
signals was at least 150 times greater than that of a model
with $k$ signals
\citep{kass:1995,tuomi:2011,tuomi:2012,tuomi:2011b}; (2) the
RV amplitudes of every signal were significantly greater than
zero \citep{tuomi:2012}; (3) and that the periods of each
signal were well-constrained from above and below because if
this was not the case, we could not tell whether the
corresponding signals were indeed of Keplerian nature and
periodic ones. These detection criteria have been used in
\citet{tuomi:2012} and they appear to provide reliable results
in terms of the most trustworthy number of signals in an RV
data set. We claim a detection of a Keplerian signal in the
data if the Markov chains of several samplings converge to a
solution that satisfies the criteria 1-3 above.

The prior probability densities were chosen to have the same
quantitative forms as in \citet{tuomi:2012}, in which e.g. the
parameter space of the RV amplitude was limited to [0, 20]
ms$^{-1}$. However, because the RV data contain the obvious
Keplerian signal of a massive candidate and a long-period
trend reported by \citet{discovery} with amplitudes clearly
larger than 20 \ms, the first two signals were allowed to
explore a wider range of semi-major amplitudes, i.e., [0, 200]
ms$^{-1}$. Also, following \citep{tuomi:2012}, we did not set
the prior probabilities of different models equal but set them
such that for models $\mathcal{M}_{k}$ and
$\mathcal{M}_{k+1}$, it holds that the prior probabilities
satisfy $P(\mathcal{M}_{k}) = 2P(\mathcal{M}_{k+1})$ for all
values of $k$.

\section{Stellar properties, observations and previous work}\label{sec:star}

\begin{table}

\caption{Basic parameters of GJ 676A}
\label{tab:star}
\centering                                      
\begin{tabular}{lrr}
Parameter               & Value                & Reference \\
\hline\hline\\
R.A.                    &  17 30 11.203        & (a) \\
Dec.                    & -51 38 13.104        & (a) \\
$\mu^*_{\rm R.A.}$
$[$mas yr$^{-1}]$       & -260.02 $\pm$  1.34  & (a) \\
$\mu_{\rm Dec}$
[mas yr$^{-1}$]         & -184.29 $\pm$  0.82  & (a) \\
Parallax [mas]          &   60.79 $\pm$  1.62  & (a) \\
V$^a$                   & 9.585   $\pm$  0.01  & (b) \\
K$^b$                   & 5.825   $\pm$  0.03  & (c) \\
Sp. type$^{c}$          & M0V                  & (b) \\  
Mass [M$_\odot$]$^d$    & 0.71    $\pm$ 0.04   & (d) \\    
Fe/H                    & +0.23   $\pm$ 0.10   & (e) \\    
Mean S-index            & 1.40    $\pm$ 0.01   & (f) \\    
\end{tabular}
\tablefoot{
(a) HIPPARCOS catalogue, \citep{hipparcos}
(b) \citep{photometry}
(c) 2MASS catalogue, \citep{twomass}
(d) Using \citep{delfosse:2000}
(e) Using \citep{johnson:2009}
(f) This work
}
\end{table}

GJ 676 is a common proper-motion pair of M dwarfs. The primary
(GJ 676A) has been classified as an M0V star
\citep{photometry}. Using the empirical relations of
\citet{delfosse:2000}, the 2MASS JHK photometry
\citep{twomass}, and its trigonometric parallax
\citep{hipparcos}, we derive a mass of 0.71 $M_\odot$ for GJ
676A. The star does not show strong evidence of activity or youth 
and, therefore, it is a good candidate
for high-precision RV studies\citep{discovery}. The basic
parameters of GJ 676A are given in Table \ref{tab:star}. The
fainter member of the pair (GJ 676B) has been classified as an
M3V and is currently separated 50 $\arcsec$ from A. From
its HIPPARCOS parallax, this corresponds to a minimum
separation of $800$ AU and an orbital period longer than 20
000 years. At this separation, the maximum acceleration of GJ
676A caused by GJ 676B on our line of sight is about $0.05$ m
s$^{-1}$ yr$^{-1}$.

New radial velocity measurements were obtained using the
HARPS-TERRA software \citep{anglada:2012a} from HARPS spectra
recently made public through the ESO archive. The spectra are
provided extracted and wavelength calibrated by the HARPS-DRS. 
Each HARPS spectrum consists of 72
echelle appertures covering the visible spectrum between 3800
and 6800 \AA. The average spectral resolution is
$\lambda/\delta\lambda = 110 000$ and each echelle apperture
consists of 4096 extracted elements (or pixels). The set of
public 75 spectra have been obtained by several programmes over
the years and typical exposure times vary between 300 to 900
seconds. The mean signal-to-noise ratio (S/N) at 6000 \AA\ is
60 and, in a few cases, it can be as low as 22. Doppler
measurements derived with HARPS-TERRA are differential against
a very high S/N template spectrum generated by coadding all
observations. The secular acceleration effect
\citep{zechmeister:2009} was subtracted from the RVs using the
HIPPARCOS \citep{hipparcos} proper motion and parallax of the
star.

One (probably two) sub-stellar companions were already reported for the
system in \citet{discovery}. The most prominent one is a massive
gas giant candidate with a period of $\sim$ 1060 days and a
semi-major amplitude of $\sim$ 120 \ms. Strong evidence for a
second, very long period candidate was also proposed by
\citet{discovery} because of a strong trend detected in the
residual to the one-planet fit. \citet{discovery} already noted
that the magnitude of this trend ($\sim$ 8 \ms yr$^{-1}$) was
too high to be explained by the gravitational pull of GJ 676B
(max value of  $\sim$ 0.05 m s$^{-1}$). Even after subtracting a
model with one planet and a trend, \citet{discovery} also noted
that the RMS of the residuals was significantly higher
($\sim$3.6 m s$^{-1}$) than the reported uncertainties (1 to 1.5
\ms), which was suggestive of potential additional candidates.
A reanalysis of the 40 spectra available to \citet{anglada:2012a}
confirmed that, even with the increased precision derived using
HARPS-TERRA (RMS 3.2 m s$^{-1}$), the star did show a
significant excess of RV variability. 

In a preliminary analysis of the new 75 HARPS-TERRA RVs, 
periodograms of the residuals to the two Keplerian solution
(gas giant + trend) showed several tentative high peaks at
36, 59, and 3.6 days. While a solution including the 36 and
3.6-day signals provided a very extreme reduction of the RMS
(from 3.1 to 1.6 \ms), the peaks in the periodograms of the
residuals provided analytic FAP estimates too high to be
acceptable ($\sim 5\%$). A preliminary Bayesian analysis of
the same new RVs \citep[methods described in ][]{tuomi:2012}, 
also indicated that additional candidates were strongly 
favoured by the data. As we will show in the analysis 
section, the RV measurements of GJ 676A are a textbook 
example where signal correlation
prevents the detection of lower amplitude signals using
periodogram methods based on the analysis of the residuals
only.

\section{Planetary system : new candidates} \label{sec:analysis}

\subsection{Recursive periodogram analysis}

\begin{figure}
\center
\includegraphics[width=0.4\textwidth, clip]{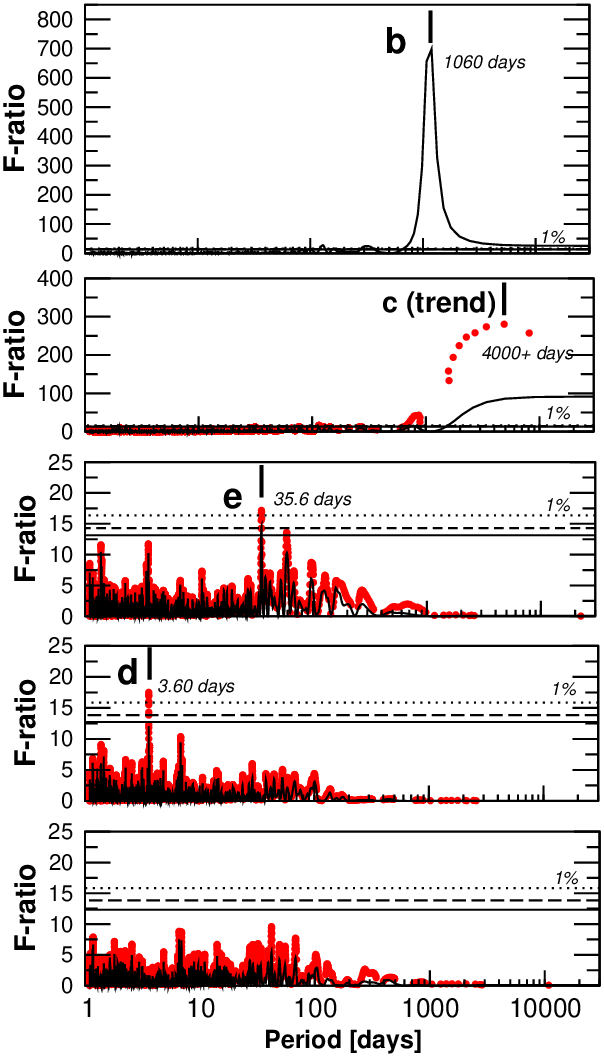}

\caption{Detection periodograms from most significant signal to
less significant one (top to bottom). Black lines are
least-squares periodograms computed on the residuals to the
k-planet model. The red dots represent the refined orbital
solution with k+1-planets at each test period as obtained 
by the recursive periodogram. The resulting sampling of 
the red dots is not uniform in frequency because the tested 
k+1 period is also allowed to adjust.}
\label{fig:periodograms}

\end{figure}

For the recursive periodogram analysis and FAP computations, 
a 1.0 m s$^{-1}$ jitter was added in quadrature to the nominal uncertainty of
each RV measurement. This value was chosen because, for
any multi-planet solution we attempted, about $\sim$ 1.0
\ms always had to be added in quadrature to match the
nominal uncertainties to the RMS of the residuals. As a
double check of the robustness of the solution, we
repeated the analysis assuming a jitter level of 0.5
\ms, 1.5 \ms, 2.0 \ms and 2.5 \ms. The 0.5 \ms value 
is the minumum uncertainty
that, according to \citet{bonfils:2011}, has to be added
to each measurement to acccount for the uncertainties in
the wavelength solution and intra-night stability of
HARPS, while 2.5 \ms would correspond to the random jitter on
a moderately active M dwarf \citep[e.g. \object{GJ 433} 
and \object{HIP 12961}
announced in ][ respectively]{delfosse:2012,discovery}. The
results obtained using different jitter assumptions were
slightly different, but still produced
the same four-planet solution. These alternative searches
are briefly discussed at the end of the section.

\begin{figure*}
\center
\includegraphics[clip,width=0.5\textwidth]{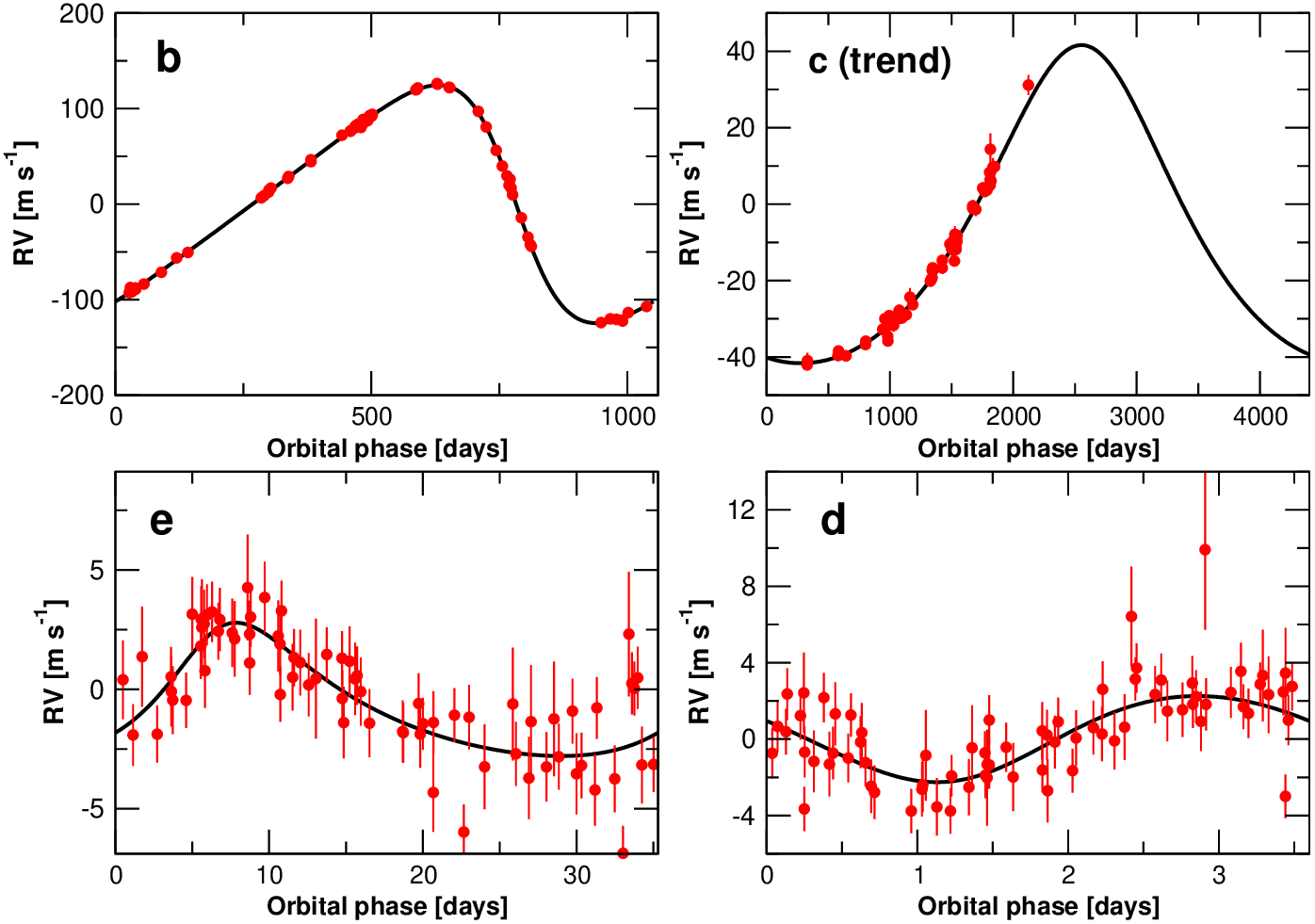}

\caption{Phase-folded radial velocity curves of the reported
new planet candidates. Even though curvature is clearly
detected (top right panel), the orbit of the of longer period
companion is still poorly constrained.
}\label{fig:phase_folded}

\end{figure*}

As seen in the top panel of Figure \ref{fig:periodograms},
there is little doubt on the reality of the first previously
reported candidate GJ676Ab \citep{discovery}. As a second
signal and instead of fitting a trend, 
we performed a recursive periodogram search
for a second planet with periods between 1.1 and 50 000 days,
obtaining a preferred solution of about 4000 days. As for 
GJ 676Ab, there is little
doubt on the statistical significance of this signal/trend
(analytic FAP threshold of 1\% is around 15, while the signal
has an F-ratio of several hundreds), and a peak 
at \textit{only} twice the time baseline indicates
the detection of significant curvature (see top-left panel in
\ref{fig:phase_folded}). As shown in the second
panel of Figure \ref{fig:periodograms}, the recursive
periodogram (red dots) compared to the periodogram of the
residuals (black line) is able to massively improve the
significance of this second signal thanks to the simultaneous
adjustment of the orbit of the first candidate. As discussed
in the Bayesian analysis section, the period and parameters
of this candidate are poorly constrained and only some
nominal values are given for reference. For detection
purposes only, we conservately assume that it can be adequately
reproduced by a full Keplerian solution and added it to the
model.

After the first two signals were included, the recursive
periodogram search for a third companion revealed one
additional periodicity at $\sim$ 35.5 days (F-ratio$\sim$
17.5). The analytic FAP was 0.155 \%, which warranted the
empirical FAP computation. In the first 1000 trials, five trials
generated false alarms (FAP $\sim0.5\%$), meaning that more
trials were necessary to securely asses if the FAP is $<1\%$.
An extended run with 10$^4$ trials produced an empirical FAP
of $\sim 0.44\%$, therefore the candidate was finally accepted. These
candidate (GJ 676Ae) corresponded to a super-Earth/sub-Neptune
mass candidate with $M \sin i\sim 11 M_\oplus$. Even though
the preferred eccentricity was rather high ($\sim$ 0.6)
quasi-circular orbits are still allowed by the data. This
candidate would receive $\sim$2.6 times more stellar radiation
than the Earth receives form the Sun. According to \citet{selsis:2007}, 
it means it would hardly be able to keep liquid water on its surface. 

Again, this 35.5 day candidate was included in the models as
a third full Keplerian signal and a recursive periodogram
search was obtained to look for additional companions. A strong
isolated peak (F-ratio 19.5, P=3.6 days) was the next
promising signal, showing an analytic FAP as low as 0.15
\%. Only one test over the first 1000 trials generated an
spurious peak with higher power, indicating that the FAP is
significantly lower than 1\%, and the candidate was immediately
accepted. The new candidate (GJ 676Ad) has a minimum mass
of $\sim$4.5 $M_\oplus$ and it is certainly too close to the
star to support liquid water on its surface. 

The recursive periodogram search for a fifth signal showed that
the next tentative periodicities have analytic FAP at the
10\% or higher level, which did not satisfy our preliminary
detection criteria and accordingly we stopped searching for additional
candidates. Even though four planet signals might seem a lot
given that \textit{only 75} RVs were used,
the amplitudes of the close-in low mass companions are
relatively high ($2-3$ \ms, see Figure
\ref{fig:phase_folded}) compared to the final RMS of the
solution (1.6 \ms) and the nominal uncertainties.

As discussed at the begining of this Section, we tested the 
robustness of the four-planet solution by applying the recursive periodogram
approach assuming different levels of jitter. Table
\ref{tab:jitter} lists the analytic FAP estimates obtained
using different jitter levels. Table \ref{tab:jitter} shows that
the analytic FAP for the fourth candidate becomes
even lower when higher jitter levels are assumed. This is,
the signals become more significant when the RV
measurements are given more similar weight, which is
equivalent to admitting that a significant contribution to the
noise (stellar and/or instrumental) is not accounted 
for in the individual measurement uncertainties. In
the next section, we show that once a converging solution is
found, a fully Bayesian approach can consistently account for
the unknown ammount of jitter and still identify the same
four candidates as the most likely periodicities in the data.

\begin{table}
\caption{Analytic false-alarm probability of planet candidates
e and d as a function of assumed stellar jitter.}
\label{tab:jitter}
\centering                                   
\begin{tabular}{lcc}
Jitter & FAP e & FAP d  \\
(\ms)  & (\%)  & (\%)   \\
\hline\hline
0.5    &  0.15 &  0.42  \\
1.0    &  0.16 &  0.16  \\
1.5    &  0.11 &  0.08  \\
2.0    &  0.16 &  0.06  \\
\hline\hline
\end{tabular}
\end{table}

\subsection{Bayesian analysis}

As discussed, there no doubt that RV data of GJ 676A
contain the signal of a massive planet ($m_{p} = 4.9$M$_{\rm
Jup}$) with an orbital period of roughly 1060 days
\citep{discovery} and a long period trend. A model with a
Keplerian signal and a linear trend was chosen as the starting
point of the Bayesian analyses.

While spotting the signature of the massive planet in the RVs
was trivial, we observed that instead of a linear trend, the
samplings preferred a second Keplerian, indicating 
significant curvature. Therefore, the second model
to be tested contains the trend modelled as a Keplerian.
However, because the long-period signal could not be
constrained, we fixed its eccentricity and period to their
most probable values in the parameter space (for period, this
space was the interval between 1 and $10T_{obs}$ days, where
$T_{obs}$ is the baseline of the data) throughout the
analyses. Because the orbit is only partially covered by the
time-baseline of the observations, we could not constrain its
other parameters much either, therefore only the MAP values for this
candidate are given in Table \ref{tab:parameters} as a
reference. Fixing period and eccentricity, however, allowed us
to draw representative samples from the parameter space and to
calculate reliable estimates for the Bayesian evidence of
each model. The curvature in the long-period trend was so
clearly present in the data that including curvature (through
a fixed period-eccentricity Keplerian) increased the model
probability by a factor of 1.8$\times 10^{13}$ and decreased
the RMS of the residuals from 4.59 to 3.00 ms$^{-1}$ (Table
\ref{probabilities}).

\begin{table}
\center
\caption{Relative posterior probabilities of models with $k=1, ..., 4$ Keplerian signals ($\mathcal{M}_{k}$) with or withour a linear trend (LT), the Bayesian evidences $P(d | \mathcal{M}_{k})$, and RMS values.}\label{probabilities}
\begin{tabular}{lccc}
\hline \hline
$k$ & $P(\mathcal{M}_{k} | d)$ & $\log P(d | \mathcal{M}_{k})$ & RMS [ms$^{-1}$] \\
\hline
1+LT & 1.0$\times10^{-30}$ & -224.0 & 4.59 \\
2 & 1.8$\times10^{-17}$ & -192.8 & 3.00 \\
3 & 3.6$\times10^{-12}$ & -179.9 & 2.20 \\
4 & $\sim$ 1 & -152.9 & 1.67 \\
\hline \hline
\end{tabular}
\end{table}

We continued by adding a third Keplerian signal to the
statistical model and performed samplings of the corresponding
parameter space. The Markov chains quickly converged to a
solution that contained the same periodic signal at 35.4 days
that was also spotted by the recursive periodograms. The model with
$k=3$ Keplerians was found to have a posterior probability
2.0$\times 10^{5}$ times that of a model with $k=2$ Keplerians
(Table \ref{probabilities}). The signal at 35.4 days
corresponds to a planet candidate with a minimum mass of 11.5
M$_{\oplus}$. When sampling the parameter space of a
four-Keplerian model, we identified a fourth strong signal in
the data with a period of 3.60 days. This was, again, the same
fourth period spotted by the recursive periodogram. Our solution
of the model with $k=4$ further increased the model
probability by a factor of 2.8$\times 10^{11}$ compared to a
model with $k=3$, so we could conclude that this 3.60 day
periodicity was also very confidently present in the data
(Table \ref{probabilities}).

\begin{figure*}
\center
\includegraphics[angle=-90,width=0.24\textwidth]{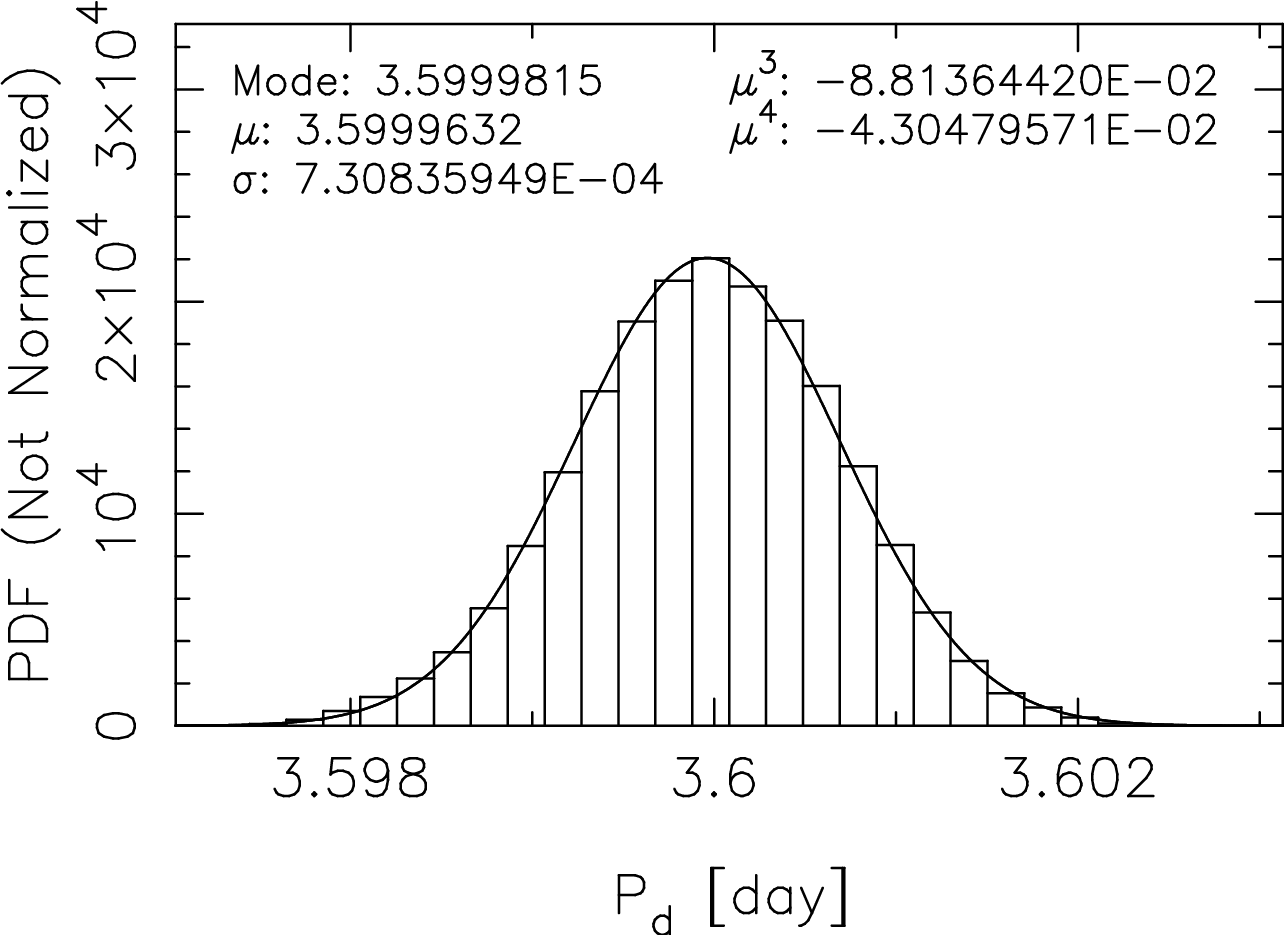}
\includegraphics[angle=-90,width=0.24\textwidth]{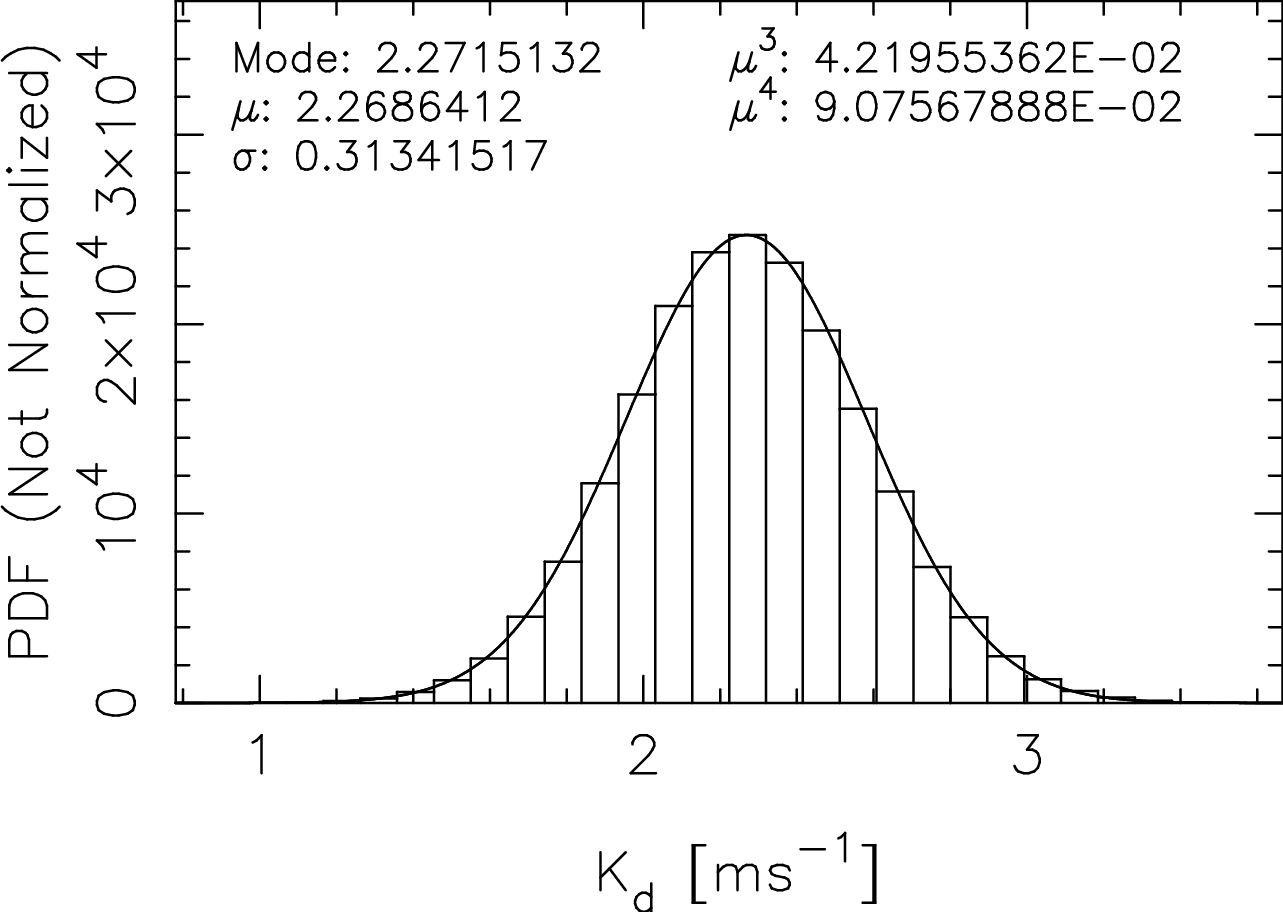}
\includegraphics[angle=-90,width=0.24\textwidth]{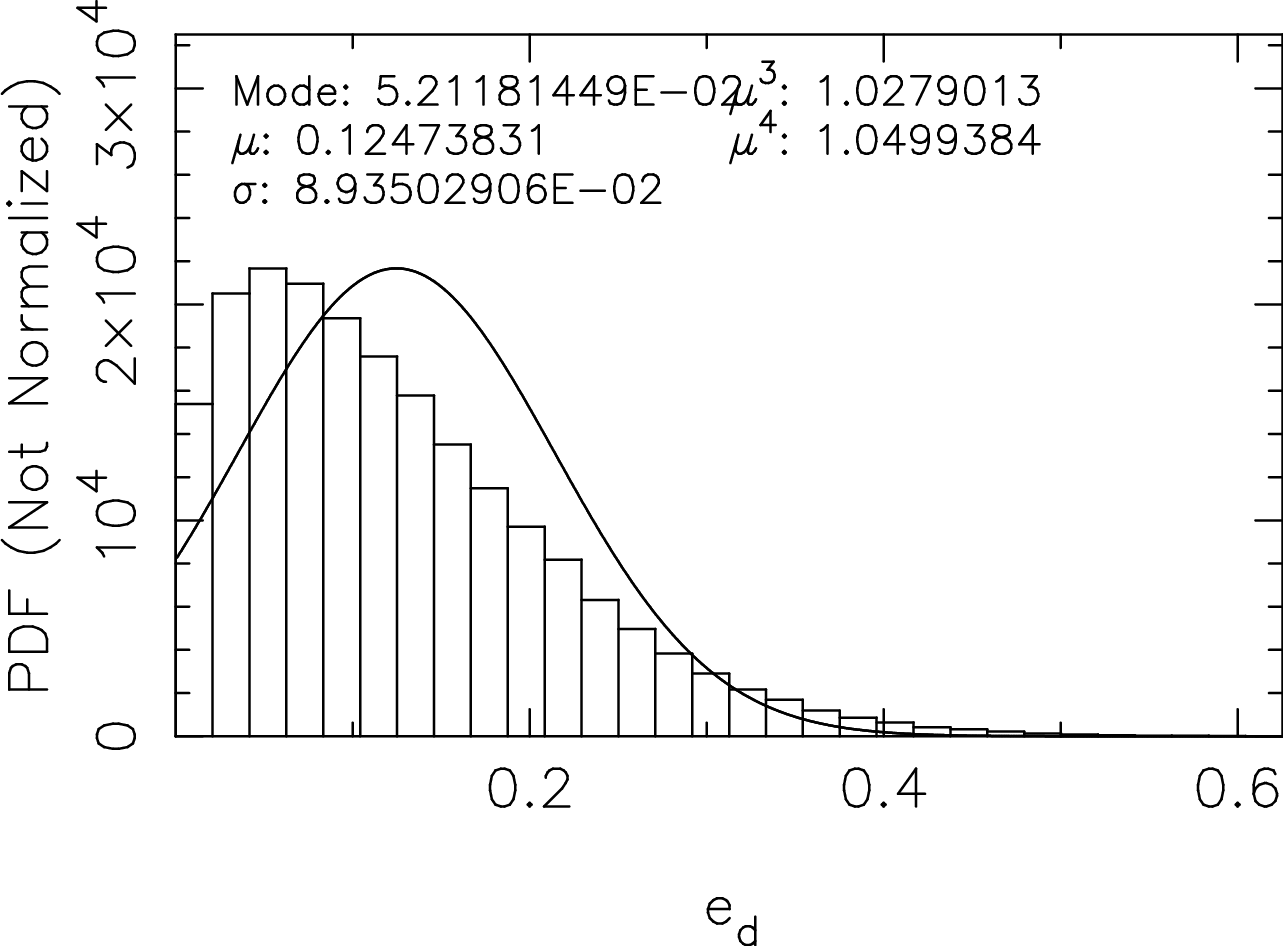}

\includegraphics[angle=-90,width=0.24\textwidth]{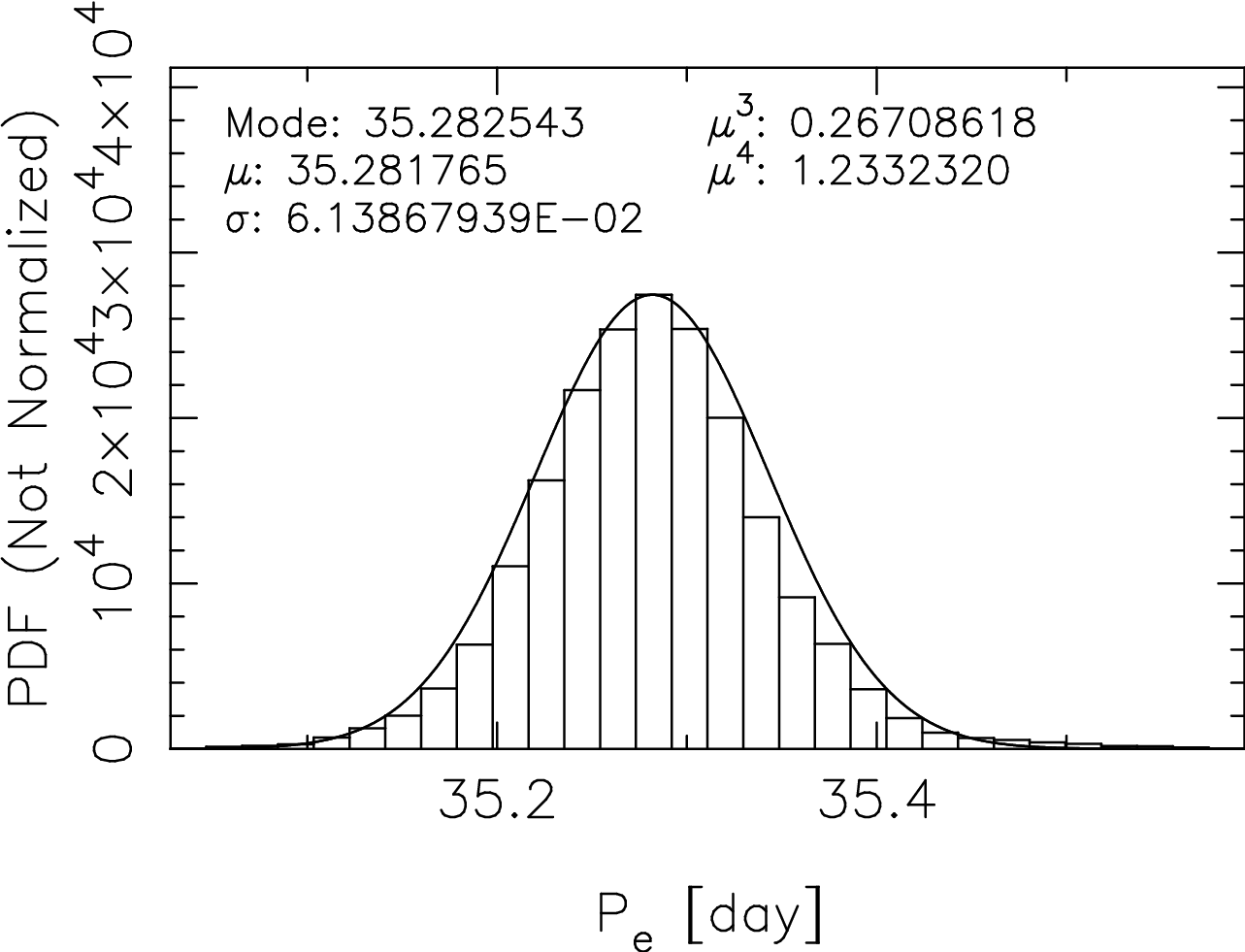}
\includegraphics[angle=-90,width=0.24\textwidth]{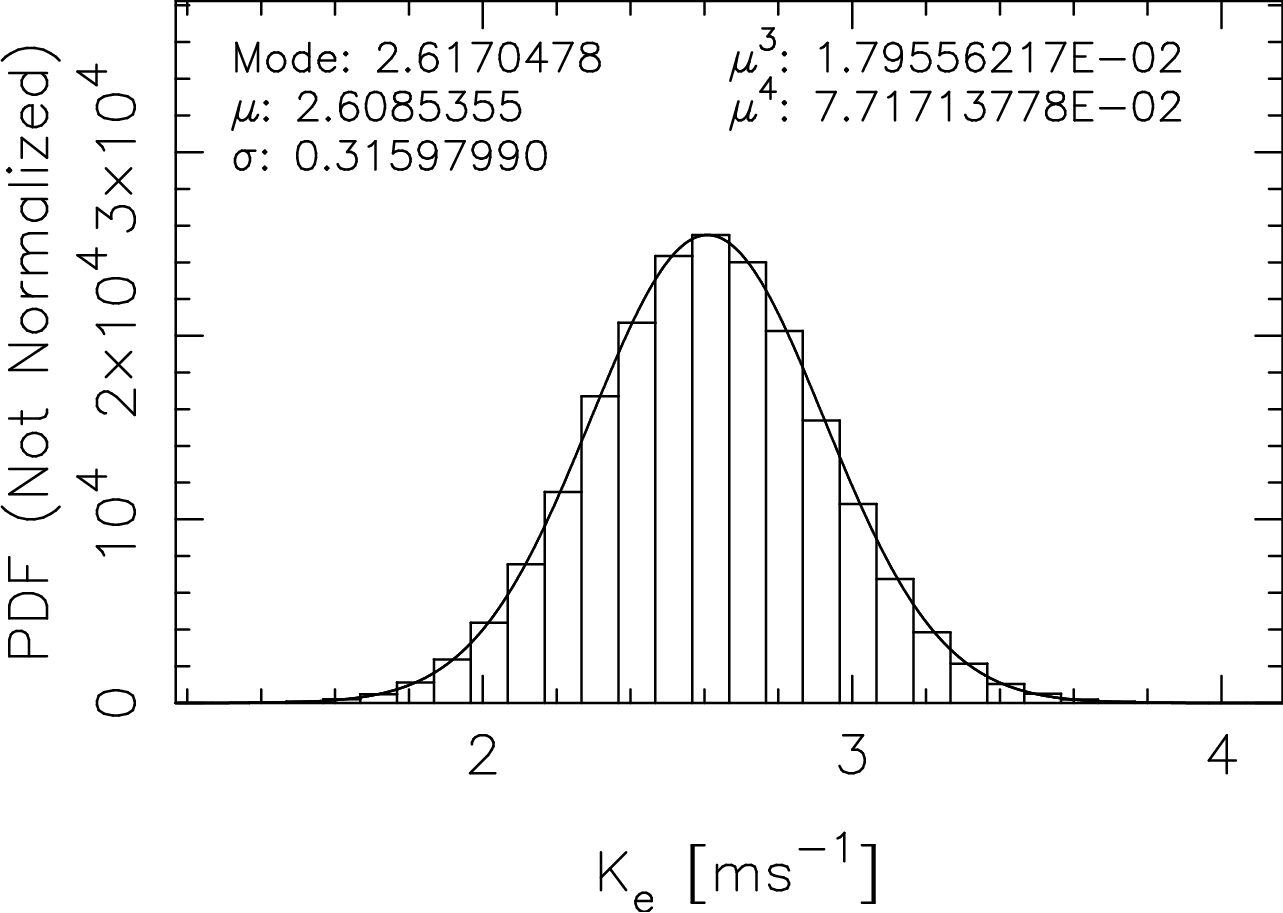}
\includegraphics[angle=-90,width=0.24\textwidth]{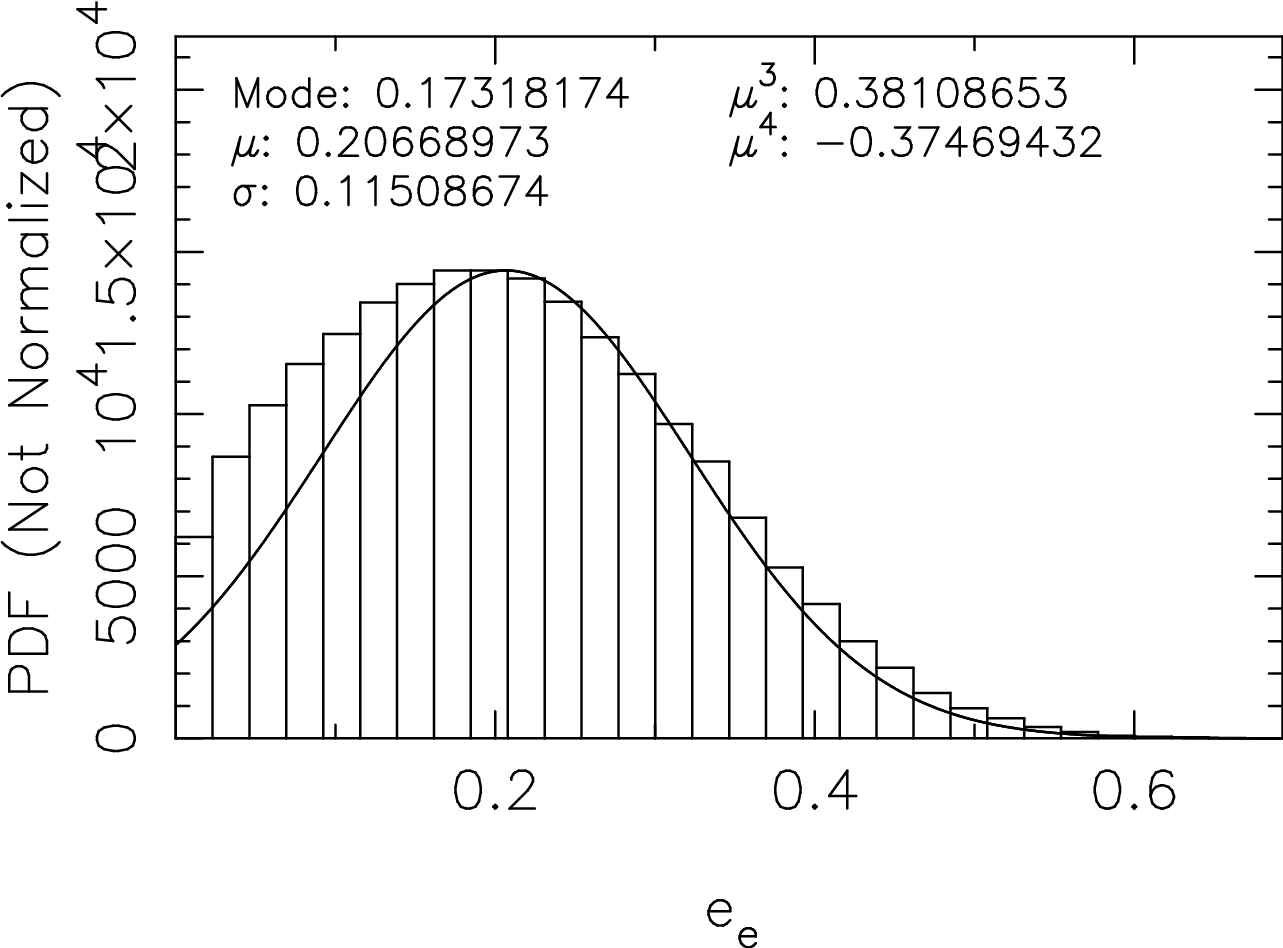}

\includegraphics[angle=-90,width=0.24\textwidth]{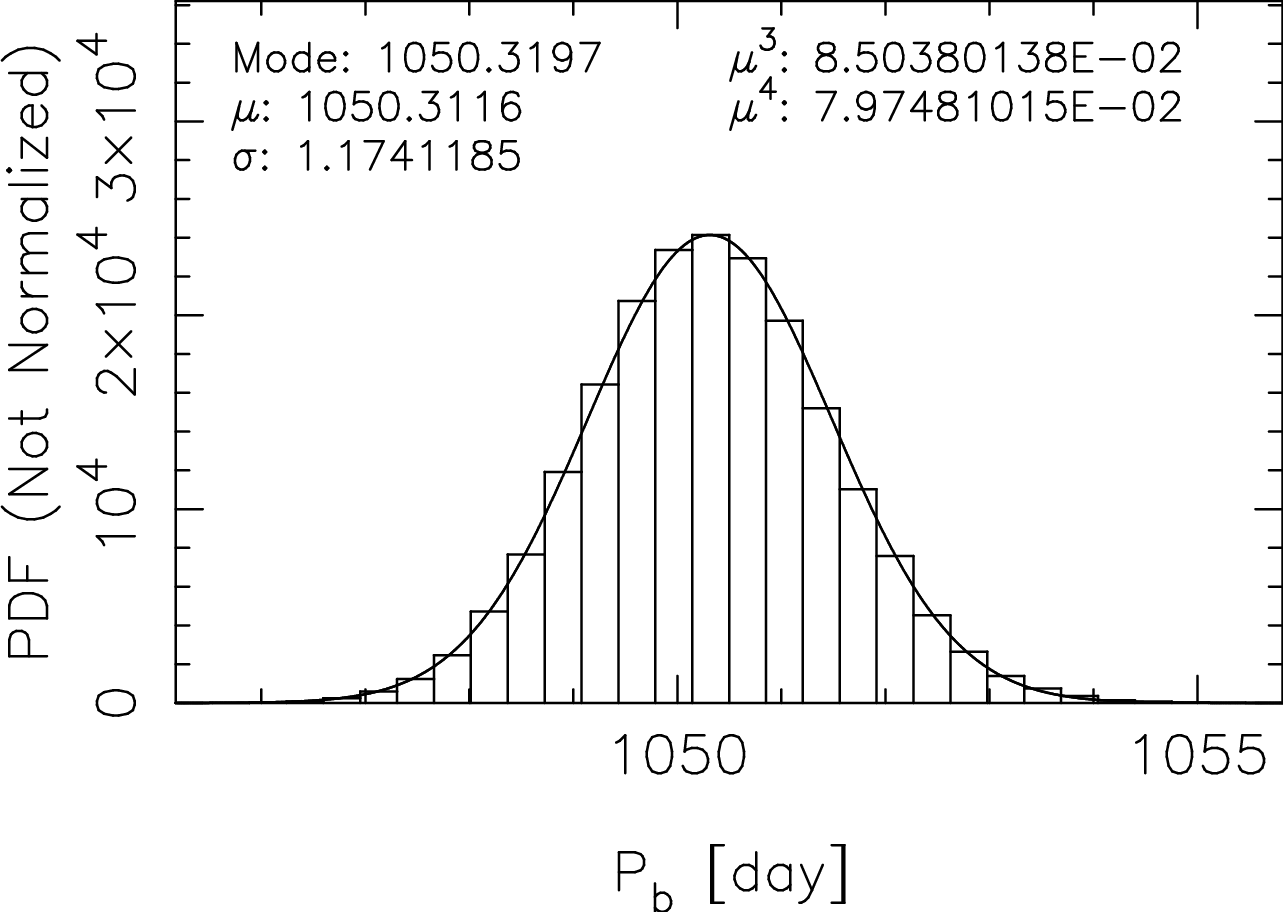}
\includegraphics[angle=-90,width=0.24\textwidth]{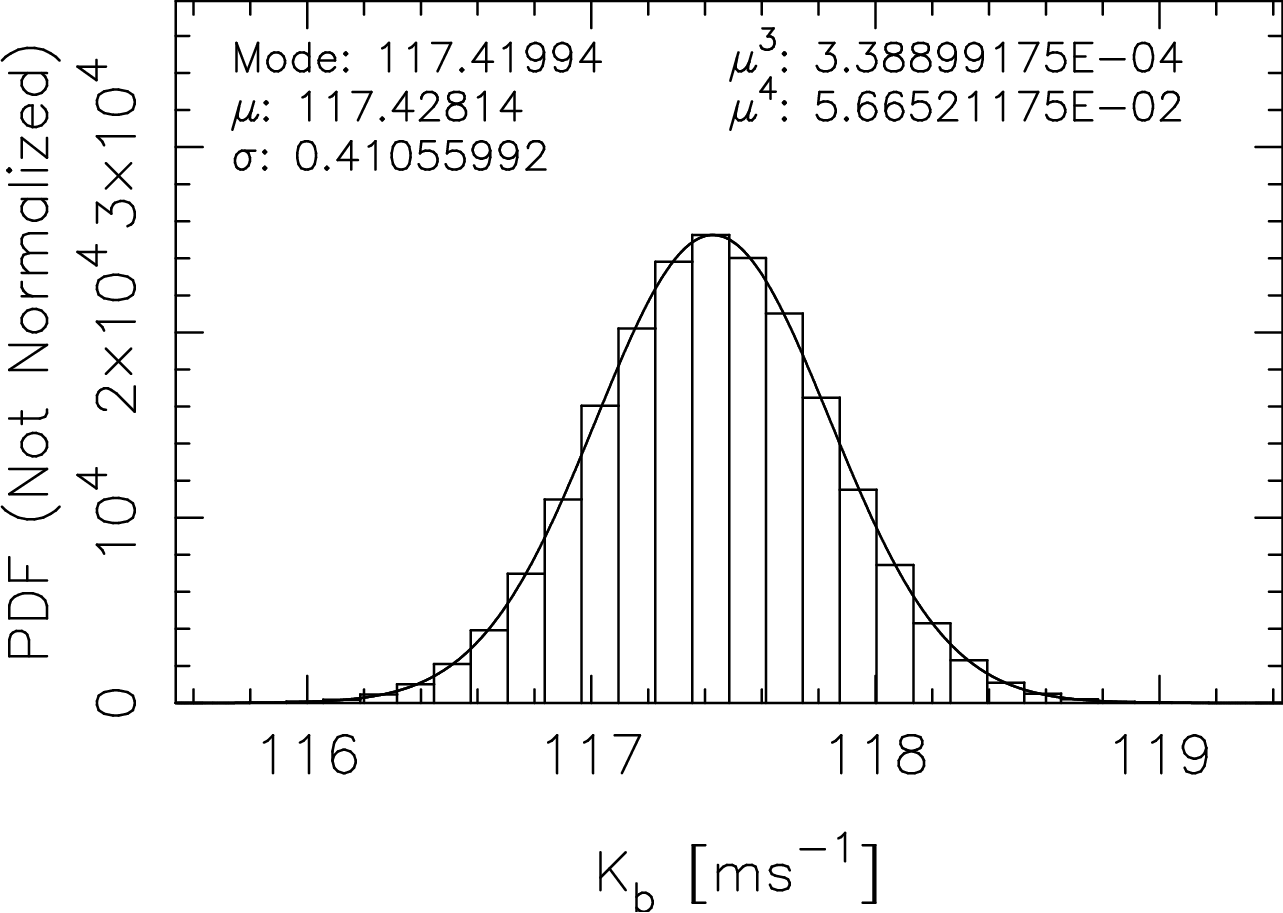}
\includegraphics[angle=-90,width=0.24\textwidth]{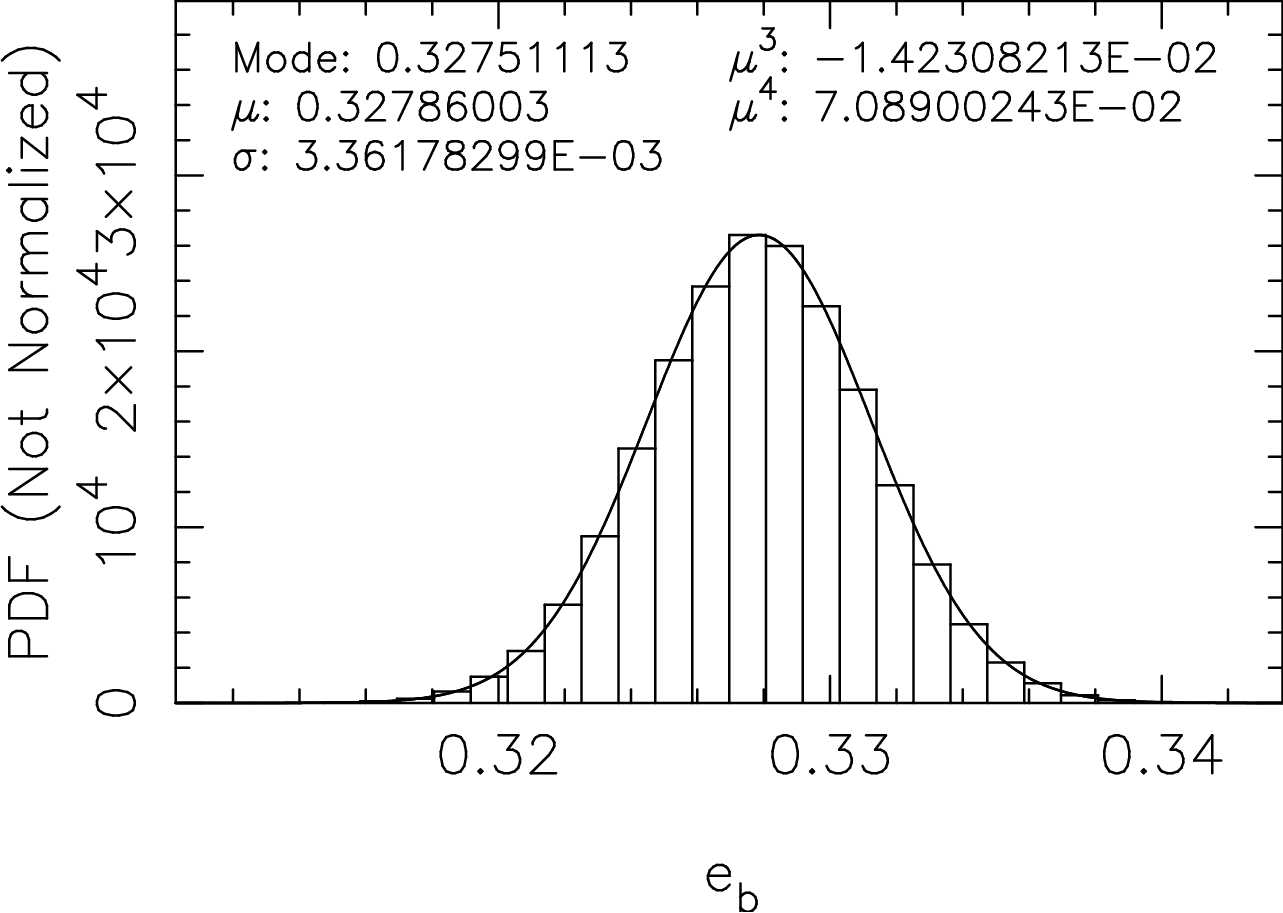}

\caption{Distributions estimating the posterior densities of
orbital periods ($P_{x}$), radial velocity amplitudes
($K_{x}$) and eccentricities ($e_{x}$), and three
constrained Keplerian signals. The solid curve is a Gaussian
density with the same mean ($\mu$) and variance
($\sigma^{2}$) as the parameter distribution. Additional
statistics, mode, skewness ($\mu^{3}$) and kurtosis
($\mu^{4}$) of the distributions are also shown.}

\label{fig:densities}

\end{figure*}

The search for additional periodic signals failed to identify
significant periodicities so we conclude that the model
probabilities imply the existence of four Keplerian signals: the
massive companion GJ 676Ab at 1.8 AU; a trend with some
curvature suggesting the presence of another massive giant
planet in a long-period orbit; and two previously unknown
planet candidates with orbital periods of 3.60 and 35.4 days and
minimum masses of 4.4 and 11.5 M$_{\oplus}$ (Table
\ref{tab:parameters}; Fig. \ref{fig:phase_folded}). These
signals satisfied all detection criteria. That is,
the radial velocity amplitudes were strictly positive
and their periods, apart from the long-period signal, were
well-constrained and had narrow distributions in the parameter
space. In addition to the MAP parameter estimates, standard
errors, and 99\% BCSs in Table \ref{tab:parameters}, we show
the distributions of the periods, RV amplitudes, and
eccentricities in Fig. \ref{fig:densities}. These
distributions show that -- apart from the eccentricities of
the two new low-mass companions, which peaked close to zero --
all densities were close to Gaussian in shape.

\begin{table*}
\center

\caption{Orbital solution of the three innermost companions
of GJ 676A and the excess RV jitter. MAP estimates, the
standard errors, and the 99\% BCSs.\label{tab:parameters}}

\begin{tabular}{lcccc}
\hline \hline
Parameter & \object{GJ 676Ad} &  \object{GJ 676Ae} &  GJ 676Ab & \object{GJ 676Ac}(trend)\tablefootmark{*} \\
\hline
$P$ [days] & 3.6000$\pm$0.0008 [3.5978, 3.6022] & 35.37$\pm$0.07 [35.10, 35.45] & 1050.3$\pm$1.2 [1046.9, 1053.7] & 4400 \\
$e$ & 0.15$\pm$0.09 [0, 0.42] & 0.24$\pm$0.12 [0, 0.56] & 0.328$\pm$0.004 [0.318, 0.339]                          & 0.2  \\
$K$ [ms$^{-1}$] & 2.30$\pm$0.32 [1.35, 3.19] & 2.62$\pm$0.32 [1.66, 3.57] & 117.42$\pm$0.42 [116.18, 118.66]      & 41  \\
$\omega$ [rad] & 5.5$\pm$1.9 [0,2$\pi$] & 5.8$\pm$2.2 [0, 2$\pi$] & 1.525$\pm$0.012 [1.491, 1.557]                & 6.21 \\
$M_{0}$ [rad]& 4.1$\pm$1.7 [0, 2$\pi$] & 0.9$\pm$2.0 [0, 2$\pi$] & 0.957$\pm$0.036 [0.844, 1.056]                 & 3.1   \\
$\sigma_{j}$ [ms$^{-1}$] & 1.38$\pm$0.18 [0.95, 1.97] \\
\,& \\
\multicolumn{2}{l}{Derived parameters}\\
\hline\hline
$a$ [AU] & 0.0413$\pm$0.0014 [0.037, 0.045] & 0.187$\pm$0.007 [0.17, 0.21] & 1.80$\pm$0.07 [1.62, 1.99]       & 5.2 \\
$m_{p} \sin i$ [M$_{\oplus}$] & 4.4$\pm$0.7 [2.4, 6.4] & 11.5$\pm$1.5 [6.5, 15.1] & 1570$\pm$100 [1190, 1770] & 951 \\
$m_{p} \sin i$ [M$_{jup}$]    & 0.014$\pm$0.002 & 0.036$\pm$0.005 & 4.95$\pm$0.31 & 3.0 \\
S/S$_0^\dagger$      &   48.1        &   2.3         &  0.025               &  0.003                \\
\hline \hline
\end{tabular}
\tablefoot{
\tablefoottext{*}{Since all parameters are poorly constrained, the 
the MAP solution is provided for orientative purposes only.}\\
\tablefoottext{\dagger}{Stellar irradiance S at the planet's orbit divided by
the flux received by the Earth from the Sun (S$_0$).}
}
\end{table*}

\subsection{Robustness of the Bayesian solution}

To assess the reliability of our solution to the GJ 676A RVs,
and indeed that of the Bayesian methods in general in
assessing the existence of Keplerian signals in RV data, we
performed a test analysis of the first 50 epochs only. The
purpose of this test was to investigate whether we could spot
the same signals, and receive the same solution from a smaller
number of observations. The 50 first epochs have a baseline of
approximately 1199 days (roughly two thirds of the full
baseline of 1794 days), and because of their lower number, we
expected them to constrain the model parameters less, i.e.
yielding broader posterior densities, and that the model
probabilities are less strongly in favour of -- possibly even
against -- the existence of the two new planet candidates
reported in this work.

Again, we started with a model containing a single Keplerian
signal and a linear trend. These were easy to spot from the
partial RV set and we could identify the same massive planet
candidate and trend reported by \citet[][69 CCF measurements
were used in that work]{discovery}. However, when we sampled
the parameter space of a two-Keplerian model, we rapidly
discovered another Keplerian signal at a 35.5 day period. The
corresponding two-Keplerian solution together with the linear
trend increased the posterior probability of the model by a
factor of 1.0$\times 10^{4}$, which clearly exceeded the
detection threshold of 150. Furthermore, we also identified a
third periodicity at 3.60 days when increasing the complexity
of the statistical model by adding another Keplerian signal to
it. This model was 5.0$\times 10^{5}$ times more probable than
the model with $k=2$, so we could conclude that three planet
candidates and a linear trend were already strongly suggested
by these initial 50 RVs. Moreover, the two new low-amplitude
periodic signals satisfy our detection criteria by having
amplitudes strictly above zero (2.27 [1.00, 3.41] ms$^{-1}$
and 2.83 [1.48, 4.04] ms$^{-1}$ for GJ 676A e and d,
respectively) and well-constrained orbital periods (3.6000
[3.5963, 3.6027] days and 35.48 [35.16, 35.90] days,
respectively). This solution is consistent with the one
received for the full data set in Table \ref{tab:parameters},
which implies that the two new planets could already have been
detected in the HARPS RVs when the 50th spectrum was obtained
back in October 2009, possibly even earlier. 

We performed the recursive periodogram analysis of the same 50
epochs. Again, the massive GJ 676Ab and the trend were also
trivially detected. Then we attempted a recursive periodogram search
for a third Keplerian. This search spotted the 35.5 day
signal as the next most likely periodicity, but provided an
analytic FAP of 15 \%, which did not satisfy our preliminary
detection criteria (analytic FAP $<1\%$). In order to check if
the 3.6 days candidate could be inferred by periodogram
methods, we added the 35.5 days signal to the model and
performed a recursive periodogram search for a fourth
candidate. Although a peak at 3.6 days was present, it
was not, by far, the most significant periodicity suggested as the
fourth signal.

This result implies that Bayesian methods are clearly more
sensitive in detecting low-amplitude signals compared to
classic periodogram approaches (even compared to our newly
developed recursive periodogram method). Even if a reasearcher
prefers to obtain frequentist confirmation (e.g., empirical
FAP) of a signal before announcing it, early Bayesian
detections can be used to optimize observational strategies
and sample the periods of interest. We are conducting
simulations with synthetic dataset to identify failure modes
of the proposed Bayesian methods (e.g., identify situations
that could generate false positives) and refine the detection
criteria accordingly.

\section{Analysis of three activity indices}\label{sec:activity}

In this section we analyse the variability of some
representative activity indices and discuss their possible
relation to Doppler signals. HARPS-TERRA obtains the CaII H+K
activity index \citep[S-index in the Mount Wilson
system, ][]{baliunas:1995} and collects the measurements provided
by the HARPS-DRS for two of the cross-correlation function (CCF)
parameters that are also sensitive to stellar activity :
bisector span (or BIS), full-width at half-maximum of the CCF
(or FWHM). 

The S-index is directly measured by HARPS-TERRA on the
blaze-corrected spectra using the definitions given by
\citet{lovis:2011} and is an indirect measurement of the
chromospheric emission. Because the strength of the magnetic
field affects the efficiency of convection, some spurious RV
signals could correlate with variability in the S-index
\citep{lovis:2011}. Magnetically active regions can also
introduce periodicities in the S-index as the star rotates
\citep[e.g][]{gj674}. The BIS is a measure of the asymmetry of the
average spectral line and should correlate with the RV if the
observed offsets are caused by spots or plages rotating with the
star \citep{queloz:2001}. The FWHM is a measure of the width of
the mean spectral line and its variability is usually associated
with changes in the convective patterns on the stellar surface
that might also induce spurious RV offsets. Since the connection
between activity and RV jitter on M dwarfs is still only poorly
understood \citep{lovis:2011}, we restrict our analyses to
evaluate if any of the indices has periodicities similar to the
detected RV candidates. 

\begin{figure}
\center
\includegraphics[width=0.45\textwidth, clip]{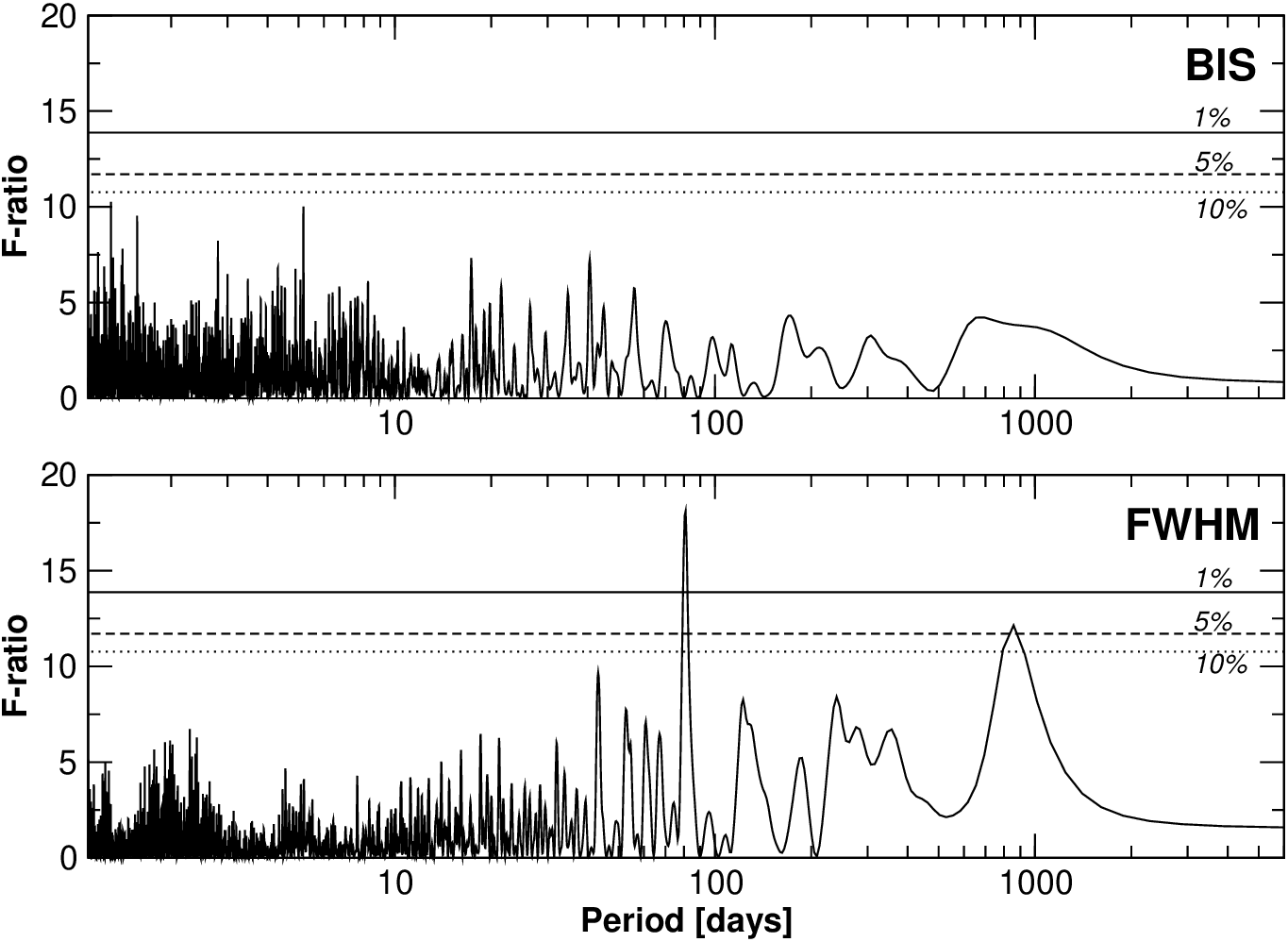}
\caption{Periodograms of the bisector span (top) and the
full-width-at-half-maximum of the cross-correlation function
(bottom). No periodic signals with analytic FAP smaller 
than 5\% were detected in either index.}
\label{fig:ccfindex}

\end{figure}

As shown in Figure \ref{fig:ccfindex} (upper panel), no
strong periodicities were detected on the BIS. However,
diagnostics based on the line symmetries have low discriminating
power for M dwarfs. A comparison active star with a similar
spectral type is AD Leo (\object{GJ 388}), which is a fast rotator
(P$\sim$2.2 days) and is magnetically active \citep{morin:2008}.
On AD Leo, BIS has been found to strongly correlate with RVs
\citep{bonfils:2011}. The amplitude of the variability of BIS
was found to be 10 times smaller \citep[$\sim$ 2
\ms][]{reiners:2012} compared to the corresponding Doppler
counterpart ($\sim$ 30 \ms). Following the same approximate
rule, and given that the RMS of BIS on GJ 676A is 4.7 \ms, only
spurious Doppler signals substantially stronger than K$\sim$5 \ms are
expected to produce any measureable effect on the BIS. The two newly
proposed candidates have amplitudes smaller than 3 \ms and,
therefore the absence of periodicities in BIS does not provide a
good diagnostic to assess the reality of these signals.

On M dwarfs, FWHM has been shown to be more effective in
identifying activity induced Doppler signals. For
example, FWHM measurements on \object{GJ 674} \citep{gj674} revealed that 
the second
signal detected in the RVs ($\sim$ 35 days) was probably related
to the presence of a persistant dark spot. Similarly, the
H$_\alpha$ index, the S-index, and additional photometric
follow-up revealed the same periodicity (which is likely related
to the rotation period of the star). Another example is the FWHM
periodicity reported by \citet{anglada:2012b} and
\citet{delfosse:2012} on the M dwarf \object{GJ 667C}. Again, the FWHM
and the S-index both showed a signal with almost identical
period, strongly suggesting that distortions of the mean spectral
line were caused by a magnetic feature
corrotating with the star. This argument was used to cast doubts
on the reality of a candidate Doppler signal at $\sim$ 91 days
(GJ 667Cd?). Applying the recursive periodogram method to the
FWHM measurements of GJ 676A, we do detect a strong isolated
periodicity at 80.75 days with an analytic FAP of 0.043 \% that
could be related to the stellar rotation. The search for a
second signal does not reveal any peak above the 10\% analytic
FAP threshold. None of the Doppler signals appear to
be remotely related to this 80.75 day period.

\begin{figure}[tb]
\center
\includegraphics[width=0.48\textwidth, clip]{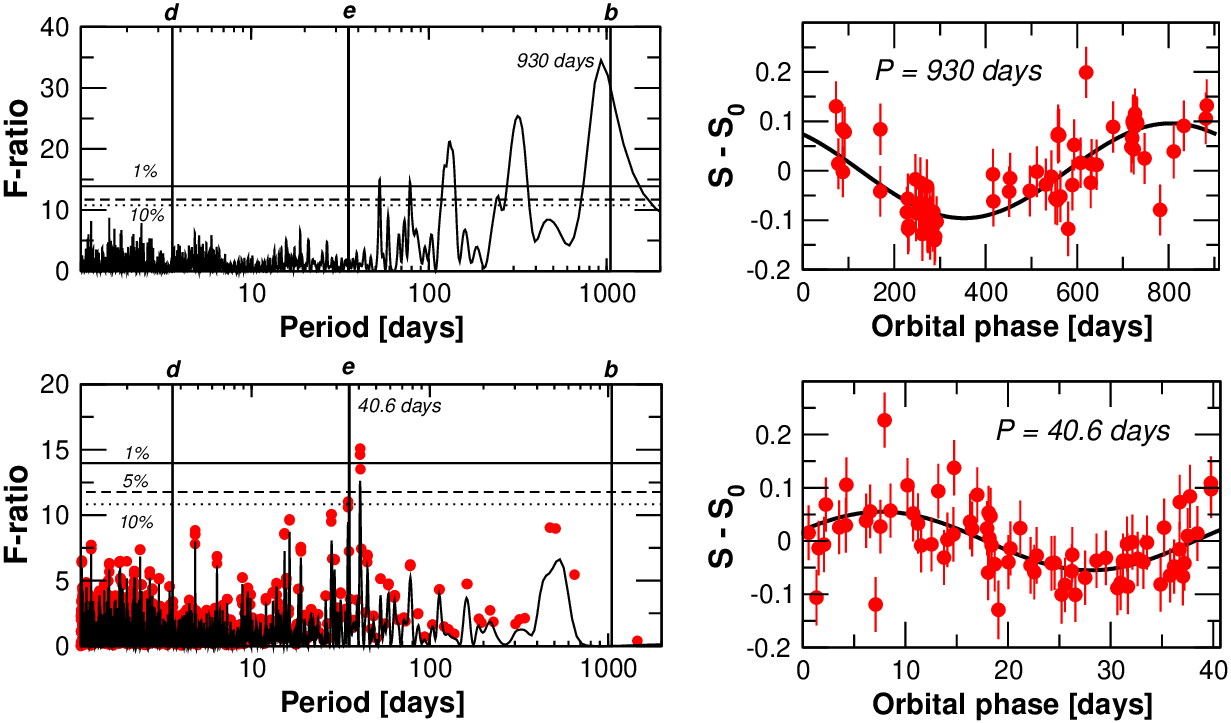}

\caption{Periodograms (left) and phase folded plots
(right) of the signals detected in the S-index. The mean
S-index value $S_0 = 1.40$ has been subtracted from the
phase folded plots for visualization purposes. The
periods of the newly proposed  planet candidate signals
are marked as black vertical bars. Their 95\% confidence
level ranges are smaller than the width of the lines.}
\label{fig:sindex}
\end{figure}

We also performed a recursive periodogram analysis of the
S-index, expecting to detect some counterpart to the FWHM
variability. Surprinsingly, the S-index does show a signal, but
with a period of $\sim$ 930 days (FAP$\sim$ 0.01\%). Although
the signal has a similar period as the GJ 676Ab candidate, this
coincidence was not mentioned in the discovery paper by
\citet{discovery}. Even if the periods are similar, two
arguments favour the Keplerian interpretation of the candidate as
a planet. First, the signal in the S-index is not in phase (or
anticorrelated) with the Doppler one. Second, \citet{silva:2012}
have shown that RV offsets correlated with the variability of the
S-index (or similar spectroscopic indices) are at the level of a
few m $s^{-1}$ while GJ 676Ab's RV semi-amplitude is about
120 \ms. Therefore, apparently the similarity in the periods
is purely coincidental. Evidence for an activity cycle of $\sim$
1000 days is also supported by the analysis of the Na I-index
performed by \citet{silva:2012} using a subset of these observations.
A search for a second signal in the S-index revealed a
second periodicity at 40.6 days (analytic FAP $\sim$ 0.028\%).
The signal is well reproduced by a sinusoid and could be related
to a magnetic feature co-rotating with the star (40.6 days is
approximately one-half of 80.7 days). While we detect a Doppler
signal at 35.4$\pm$0.07 days, this period is statistically very
distinct to 40.65 days. The recursive periodogram of the S-index
in the central panel of Figure \ref{fig:sindex} shows a weaker
peak at 34.7 days that would not qualify as a detection
(analytic FAP $\sim$ 6\%) even if it were the dominant signal in
the time-series. Moreover, this signal completely disappears when
searching for a third periodicity using the recursive
periodogram scheme (bottom panel in Figure \ref{fig:sindex}). As
a double check, we forced a sinusoids solution to this
34.7 days period and computed the recursive periodogram for a third
signal. In this case, a peak with a FAP of $\sim$ 5\% and P=40.6
days still remained, providing additional indication that 40.6 days
is indeed preferred by the S-index data. As shown in the bottom panel of
Figure \ref{fig:sindex}, and after adding the 40.6 day sinusoid, 
no signal could be identified in the
sarch for a third signal in the S-index.

In summary to the discussion of the activity, we detected
one signal in the FWHM and two periodicities in the S-index (see
Table \ref{tab:activity}). Compared to other M-dwarfs that show
indications of activity-induced periodicities (e.g., GJ 433, GJ 674,
GJ 667C), the periods found in the FWHM and the S-index do not
match, which complicates their physical interpretation. While the 
long-period signal in the S-index is likely to be caused by a
sun-like activity cycle, it is less clear if one or
the other signal is truly related to stellar rotation
\citep[both signals could match typical rotation periods
measured for M dwarfs with ages of a few Gyr, ][]{irwin:2011}. In
the absence of further diagnostics and in addition to other caveats
applicable to any Doppler candidate, we find no
reason to doubt on the Keplerian nature of the newly reported
Doppler signals.

\begin{table}
\caption{Summary of signals in activity indices.}
\label{tab:activity}
\centering                                   
\begin{tabular}{lcccc}
Index    & Analytic FAP & Period & Amplitude & Possible \\
         & (\%)         & (days) &           & Origin   \\
\hline\hline
BIS       &  -           &  none  & $<$ 3 \ms & -         \\
FWHM      &  0.043       &  80.75 & 6.9 \ms  & Rotation? \\
S-index 1 &  0.010       &  933   & 0.106    & Activity cycle \\
S-index 2 &  0.028       &  40.6  & 0.054    & Rotation? \\
\hline\hline
\end{tabular}
\end{table}

\section{Conclusions}\label{sec:conclusions}

We re-derived high-precision radial velocities of the public
HARPS observations of GJ 676A using our newly developed
HARPS-TERRA software, obtaining a significant improvement over
the RVs obtained using the CCF approach. We developed a
recursive periodogram method to enhance the sensitivity of
least-squares solvers to low-amplitude signals when strong multiplanet 
correlations are present, and we provided a recipe to
derive empirical FAPs. We compared the results obtained for the RV
of GJ 676A to the candidates identified by a Bayesian analyses, 
obtaining compatible detection of the same four signals. We
provided the favoured four planet solution together with the allowed
parameter intervals as derived from the Bayesian MCMC samplings.

While it is clear that Bayesian methods are more general and
provide a more complete description of the data, frequentists
methods (e.g., empirical FAP computations) allow a simpler
interpretation of the significance of a detection. The
combination of criteria from both approaches provides great
confidence in our results. We have shown that after the early
Bayesian detection of four planets, 25 more measurements were
sufficient to confirm the same candidates with periodogram based
methods. Even if a researcher prefers frequentist confirmation
of candidates, early Bayesian detection can be used to optimize
follow-up programmes \citep{gregory:2005}. This study shows
that the confluence of recent data analysis developments
(HARPS-TERRA, Bayesian toolbox, advanced periodograms) achieve a
significant boost in sensitivity to very low mass companions, even in
already existing datasets. Compared to the significant investment
required in hardware development, developing improved
data-analysis methods comes at a significantly lower cost, thus
enabling a more efficient utilization of the observational
resources.

GJ 676A shows indications of mild activity levels in the
form of coherent variability in the width of the mean line
profile (traced by the FWHM of the CCF) and two periodic signals
present in its chromospheric emission. However, given that none
of the signals coincides with the others, their physical
interpretation is not clear. Systematic changes of a few m/s in
the instrumental profile of HARPS have been reported by
\citet{lovis:2011}, so it is possible that the detected
variations of the FWHM have an instrumental origin.

Concerning the new two planet candidates, we find that they are
both in the sub-Neptune mass regime. The shorter period
candidate (GJ 676Ad) has a significant probability of transit
\citep[$\sim 5\%$  according to ][]{charbonneau:2007}, thus
encouraging the photometric follow-up of the star. The 
long-period companion (massive planet or brown dwarf) is
now clearly detected through significant curvature, and a 
period of $\sim$ 4000 days or longer is tentatively
suggested by the data. With \object{GJ 876} \citep{rivera:2010} and 
\object{GJ 581}\citep{mayor:2009}, GJ 676A becomes the third M dwarf with four
planet candidates detected. Except for the solar system itself, this
planetary system has the broadest range of minimum masses and
periods reported so far (from 5 M$_\oplus$ to 5 $M_{jup}$, and
from 3.6 days to 4000 or more days). Despite the abundance of
candidates, the periods (and corresponding semi-major axis) are
spaced far enough appart that we do not anticipate major dynamical
stability problems. Compared to the more dynamically packed GJ 581
and GJ 876 systems, the orbits of the candidate planets leave
ample room to detect more candidates in
intermediate orbits whenever additional RV observations become
available. Owing to the proximity of GJ 676A to our Sun
($\sim$16.4 pc), the long-period, massive candidates are
attractive targets for direct imaging attempts
\citep{lagrange:2010}. Given that the make-up of stars in binary
systems should be similar, it would be very interesting to
investigate whether \object{GJ 676B} (M3.5V) has been as prolific as GJ
676A in forming all kinds of planets.

\begin{acknowledgements}

GAE is supported by the German Federal Ministry of Education
and Research under 05A11MG3. M. Tuomi is supported by RoPACS
(Rocky Planets Around Cool Stars), a Marie Curie Initial
Training Network funded by the European Commission's Seventh
Framework Programme. We are grateful for the advise, support
and useful discussions obtained from Paul Butler (DTM), Ansgar
Reiners (IAG), Mathias Zechmeister (AIG) and Hugh Jones (HU).
We thank Sandy Keiser for setting up and managing the
computing resources available at the Department of Terrestrial
Magnetism--Carnegie Institution of Washington. This work is
based on data obtained from the ESO Science Archive Facility
under request number GANGLFGGCE178541. This research has made
extensive use of the SIMBAD database, operated at CDS,
Strasbourg, France; and the NASA's Astrophysics Data System.

\end{acknowledgements}

\bibliography{biblio}

\clearpage

\longtab{2}{
\begin{longtable}{lccccccccc}
\caption{Differential HARPS-TERRA RV measurements of 
GJ 676A measured in the solar system barycenter 
reference frame. Secular acceleration was 
subtracted from the RVs. S-index and corresponding uncertainty
are given in the Mount Wilson system. CCF parameters
for each epoch as provided by the HARPS-ESO
archive (RV$_{\rm CCF}$ not corrected by secular acceleration).
}\label{tab:rvs}
\\
\hline\hline
BJD     & 
RV           & 
$\sigma_{\rm RV}$ & 
S-index & 
$\sigma_S$ & 
RV$_{\rm CCF}$  & 
$\sigma_{\rm RV-CCF}$ &
FWHM$^\dagger$ &
BIS$^\dagger$ 
\\
(days)       & 
(m s$^{-1}$) & 
(m s$^{-1}$) &
             &            
	     & 
(km s$^{-1}$) & 
(m s$^{-1}$) &
(km s$^{-1}$) & 
(m s$^{-1}$) &
\\
\hline
\endfirsthead
\caption{continued.}\\
\hline\hline
BJD     & RV           & 
$\sigma_{\rm RV}$ & 
S-index & 
$\sigma_S$ & 
RV$_{\rm CCF}$  & 
$\sigma_{\rm RV-CCF}$ &
FWHM$^\dagger$ &
BIS$^\dagger$ 
\\
(days)       & 
(m s$^{-1}$) & 
(m s$^{-1}$) &
             &            
	     & 
(km s$^{-1}$) & 
(m s$^{-1}$) &
(km s$^{-1}$) & 
(m s$^{-1}$) &
\\
\hline
\endhead
\hline\hline
\endfoot
2453917.74799  &     -49.64 & 1.07 & 1.515 &0.0147  & 0.303407$^*$ & -     &   -        &  -          \\	    
2453919.73517  &     -42.40 & 1.74 & 1.525 &0.0177  & 0.303160$^*$ & -     &   -        &  -          \\	    
2454167.89785  &      50.99 & 0.87 & 1.276 &0.0102  & 13.81054  & 0.96  &   3.3762   &  -16.56     \\	    
2454169.89585  &      47.69 & 0.74 & 1.292 &0.0095  & 13.77035  & 0.83  &   3.3751   &  -10.33     \\	    
2454171.90444  &      51.41 & 0.76 & 1.231 &0.0102  & 13.71860  & 0.94  &   3.3752   &  -10.94     \\	    
2454232.81801  &      50.43 & 0.82 & 1.347 &0.0100  & 13.69687  & 0.78  &   3.3742   &   -5.25     \\	    
2454391.49180  &    -112.11 & 0.83 & 1.390 &0.0098  & 13.76543  & 0.76  &   3.3756   &  -13.42     \\	    
2454393.48993  &    -116.14 & 0.85 & 1.428 &0.0110  & 13.67842  & 0.87  &   3.3684   &   -8.60     \\	    
2454529.90084  &    -192.95 & 1.02 & 1.360 &0.0115  & 13.55801  & 1.03  &   3.3824   &  -17.89     \\	  
2454547.91501  &    -190.21 & 0.82 & 1.400 &0.0098  & 13.60705  & 0.79  &   3.3858   &  -12.97     \\	  
2454559.81569  &    -182.54 & 1.10 & 1.650 &0.0128  & 13.63159  & 0.98  &   3.3744   &   -7.55     \\	  
2454569.90363  &    -189.17 & 1.31 & 1.370 &0.0126  & 13.67565  & 1.19  &   3.3780   &  -11.06     \\	  
2454571.88945  &    -190.03 & 0.58 & 1.391 &0.0078  & 13.66016  & 0.56  &   3.3747   &   -9.04     \\	  
2454582.82029  &    -181.49 & 0.82 & 1.359 &0.0102  & 13.66855  & 0.82  &   3.3776   &  -16.85     \\	  
2454618.75558  &    -173.61 & 1.27 & 1.432 &0.0141  & 13.49020  & 1.33  &   3.3817   &  -12.59     \\	  
2454658.69933  &    -156.68 & 0.73 & 1.393 &0.0101  & 0.349490$^*$ & -     &   -        &  -          \\	  
2454660.66163  &    -150.04 & 0.95 & 1.407 &0.0100  & 13.44917  & 0.77  &   3.3890   &  -10.79     \\	  
2454661.77222  &    -151.70 & 0.65 & 1.440 &0.0097  & 13.48276  & 0.70  &   3.3885   &  -11.88     \\	  
2454662.67523  &    -154.15 & 1.01 & 1.445 &0.0122  & 13.42979  & 1.00  &   3.3874   &   -5.09     \\	  
2454663.81158  &    -150.58 & 0.99 & 1.389 &0.0090  & 13.49663  & 0.65  &   3.3939   &  -13.19     \\	  
2454664.79004  &    -147.60 & 1.11 & 1.440 &0.0135  & 13.41360  & 1.16  &   3.3821   &  -10.26     \\	  
2454665.78637  &    -152.21 & 0.78 & 1.445 &0.0094  & 13.41781  & 0.66  &   3.3868   &  -18.38     \\	  
2454666.69605  &    -153.22 & 0.65 & 1.476 &0.0085  & 13.43290  & 0.59  &   3.3864   &  -13.15     \\	  
2454670.67260  &    -151.17 & 1.13 & 1.489 &0.0140  & 13.44134  & 1.14  &   3.3798   &  -14.99     \\	  
2454671.60332  &    -150.03 & 1.01 & 1.494 &0.0119  & 13.45861  & 0.92  &   3.3851   &  -11.57     \\	  
2454687.56195  &    -149.86 & 0.98 & 1.454 &0.0119  & 13.49968  & 0.94  &   3.3816   &  -11.94     \\	  
2454721.55487  &    -133.47 & 1.11 & 1.372 &0.0114  & 13.69112  & 1.10  &   3.3873   &  -12.65     \\	  
2454751.49069  &    -117.35 & 2.14 & 1.426 &0.0221  & 13.44064  & 2.99  &   3.3773   &    2.50     \\	  
2454773.50237  &    -108.57 & 0.87 & 1.482 &0.0102  & 13.45558  & 0.70  &   3.3742   &  -14.55     \\	   
2454916.81980  &     -44.78 & 0.60 & 1.540 &0.0070  & 13.41987  & 0.49  &   3.3956   &   -4.20     \\	  
2454921.89297  &     -48.26 & 0.99 & 1.457 &0.0120  & 13.59319  & 1.17  &   3.4032   &   -1.35     \\	  
2454930.90684  &     -40.74 & 0.95 & 1.516 &0.0109  & 13.55822  & 0.87  &   3.3866   &   -2.65     \\	  
2454931.79510  &     -40.19 & 0.91 & 1.424 &0.0106  & 13.55113  & 0.91  &   3.3932   &  -13.97     \\	  
2454935.81778  &     -39.01 & 0.52 & 1.472 &0.0062  & 13.52808  & 0.45  &   3.3917   &  -10.45     \\	  
2455013.68661  &       1.94 & 1.01 & 1.380 &0.0110  & 13.46910  & 0.88  &   3.3840   &  -14.05     \\
2455013.74372  &       0.0  & 1.18 & 1.505 &0.0138  & 13.43115  & 1.15  &   3.3874   &  -13.81     \\
2455074.52005  &      25.37 & 0.81 & 1.251 &0.0100  & 13.70314  & 0.88  &   3.3751   &  -10.56     \\	  
2455090.50702  &      31.96 & 0.95 & 1.430 &0.0108  & 13.60774  & 0.85  &   3.3814   &   -4.02     \\	     
2455091.52880  &      30.62 & 2.30 & 1.357 &0.0267  & 13.51688  & 3.45  &   3.3837   &    5.83     \\	     
2455098.49414  &      31.37 & 0.42 & 1.327 &0.0063  & 13.74989  & 0.46  &   3.3807   &   -7.49     \\	     
2455100.54094  &      36.65 & 0.51 & 1.307 &0.0080  & 13.73708  & 0.59  &   3.3773   &  -17.27     \\	     
2455101.49047  &      33.88 & 0.91 & 1.345 &0.0127  & 13.69239  & 1.13  &   3.3789   &   -8.35     \\	     
2455102.50286  &      32.35 & 1.46 & 1.274 &0.0170  & 13.67784  & 1.91  &   3.3787   &   -3.35     \\	     
2455104.54025  &      35.68 & 0.89 & 1.277 &0.0115  & 13.77991  & 1.05  &   3.3801   &   -9.90     \\	     
2455105.52363  &      34.53 & 2.14 & 1.218 &0.0216  & 13.64133  & 2.74  &   3.3801   &  -11.19     \\	     
2455106.51997  &      35.83 & 1.05 & 1.262 &0.0107  & 13.73847  & 0.95  &   3.3753   &  -12.73     \\	     
2455111.50933  &      35.28 & 0.55 & 1.280 &0.0080  & 13.72539  & 0.60  &   3.3780   &   -8.15     \\	     
2455113.49787  &      38.06 & 0.57 & 1.277 &0.0079  & 13.83148  & 0.62  &   3.3843   &  -10.49     \\	     
2455115.51499  &      43.93 & 1.84 & 1.340 &0.0217  & 13.71039  & 2.60  &   3.3678   &  -16.52     \\	     
2455116.48753  &      38.80 & 0.64 & 1.267 &0.0076  & 13.67890  & 0.56  &   3.3679   &  -14.43     \\	     
2455117.49304  &      44.05 & 1.01 & 1.273 &0.0122  & 13.68517  & 1.08  &   3.3667   &  -13.45     \\	     
2455121.52664  &      49.34 & 1.23 & 1.328 &0.0123  & 13.57599  & 1.12  &   3.3773   &   -8.38     \\	     
2455122.50532  &      47.43 & 0.87 & 1.299 &0.0108  & 13.70427  & 0.96  &   3.3795   &   -8.36     \\	     
2455124.49783  &      46.58 & 0.56 & 1.338 &0.0068  & 13.67438  & 0.49  &   3.3776   &   -9.81     \\	     
2455127.51679  &      47.31 & 0.53 & 1.367 &0.0066  & 13.66525  & 0.47  &   3.3718   &  -10.52     \\	     
2455128.51395  &      51.17 & 0.53 & 1.338 &0.0066  & 13.72336  & 0.48  &   3.3878   &   -8.93     \\	     
2455129.49540  &      50.53 & 0.70 & 1.368 &0.0075  & 13.66345  & 0.54  &   3.3742   &   -5.76     \\	     
2455132.49575  &      50.21 & 0.74 & 1.301 &0.0081  & 13.69467  & 0.60  &   3.3741   &   -7.86     \\	     
2455133.49318  &      49.77 & 0.77 & 1.303 &0.0089  & 13.70667  & 0.68  &   3.3708   &  -10.70     \\	     
2455259.90727  &      90.78 & 1.19 & 1.396 &0.0114  & 13.70228  & 1.01  &   3.3764   &  -15.75     \\	     
2455260.86440  &      90.78 & 0.79 & 1.333 &0.0094  & 13.66861  & 0.77  &   3.3782   &   -7.05     \\	     
2455284.89313  &      84.51 & 1.37 & 1.211 &0.0142  & 13.78936  & 1.70  &   3.3662   &   -6.85     \\	     
2455340.70850  &      67.31 & 1.06 & 1.367 &0.0119  & 13.52789  & 1.03  &   3.3832   &  -11.94     \\	     
2455355.79544  &      50.55 & 0.94 & 1.348 &0.0122  & 13.67757  & 0.92  &   3.3712   &  -14.84     \\	     
2455375.61072  &      25.37 & 1.11 & 1.417 &0.0156  & 13.54259  & 1.19  &   3.3728   &   -9.21     \\	     
2455387.65668  &      10.96 & 1.39 & 1.346 &0.0179  & 13.63290  & 1.44  &   3.3669   &   -6.27     \\	     
2455396.53797  &       0.18 & 1.31 & 1.293 &0.0136  & 13.60618  & 1.24  &   3.3776   &   -9.73     \\	     
2455400.64286  &      -8.10 & 0.58 & 1.317 &0.0084  & 13.64280  & 0.60  &   3.3899   &  -11.67     \\	     
2455401.59478  &      -5.14 & 0.95 & 1.455 &0.0115  & 13.59837  & 0.88  &   3.3761   &  -11.80     \\	     
2455402.59092  &       4.03 & 4.06 & 1.467 &0.0318  & 13.61852  & 4.40  &   3.4018   &  -11.53     \\
2455404.64556  &      -9.17 & 1.97 & 1.481 &0.0197  & 13.85158  & 2.54  &   3.4019   &   -8.21     \\	   
2455407.57676  &     -16.79 & 0.97 & 1.380 &0.0146  & 13.62195  & 1.16  &   3.3842   &   -4.98     \\	   
2455424.57544  &     -40.63 & 2.18 & 1.268 &0.0229  & 13.62399  & 2.93  &   3.3805   &  -23.28     \\	   
2455437.61843  &     -56.65 & 0.77 & 1.404 &0.0133  & 13.56143  & 0.88  &   3.3743   &  -14.78     \\	   
2455711.71907  &     -94.96 & 2.41 & 1.512 &0.0243  & 13.56101  & 2.66  &   3.3921   &  -17.64     \\	   

\end{longtable}
\tablefoot{
\\
$^\dagger$ Based on \citet{pepe:2011}, formal uncertainty
in FWHM and BIS are $2.35 \sigma_{RV-CCF}$ and $2.0
\sigma_{RV-CCF}$, respectively.\\
$^*$ HARPS-ESO cross-correlation function algorithm produced
an unreliable measurement. 
None of the CCF indices on this epoch were used in the analyses.
}
}

\end{document}